\documentclass[aps,showpacs,prd,floatfix,twocolumn]{revtex4}
\usepackage{amsmath,amssymb}
\usepackage{graphicx}
\usepackage{pst-plot}
\usepackage{mathrsfs}
\usepackage{mathtools}
\usepackage{pst-func}
\usepackage{rotating}
\usepackage{epsfig}

\usepackage{chemarrow}

\usepackage[titletoc]{appendix}

\newtheorem{Joy}{Definition}[section]

\begin{document}

\title{Macroscopic Observability of Spinorial Sign Changes under ${2\pi}$ Rotations}

\author{Joy Christian}

\email{joy.christian@wolfson.ox.ac.uk}

\affiliation{Wolfson College, University of Oxford, Oxford OX2 6UD, United Kingdom}

\affiliation{Department of Materials, University of Oxford, Parks Road, Oxford OX1 3PH, United Kingdom}

\begin{abstract}
The question of observability of sign changes under ${2\pi}$ rotations is considered. It is shown that in certain circumstances
there are observable consequences of such sign changes in classical physics. A macroscopic experiment is proposed
which could in principle detect the ${4\pi}$ periodicity of rotations.
\end{abstract}


\maketitle

\parskip 5pt

\section{Introduction}

It is well known that the state of a rotating body in the physical space depends in general not only on its local
configuration but also on its topological relation to the rest of the universe \cite{Misner}. While the former feature
of space is familiar from everyday life, the latter feature too can be demonstrated by a simple rope trick, or Dirac's
belt trick \cite{Penrose}\cite{Hartung}. What is less easy to establish, however, is whether there are observable effects
of the latter relation---often referred to as ``orientation-entanglement relation'' \cite{Misner}---in the macroscopic domain
\cite{Hartung}. The purpose of the present paper is to describe a macroscopic experiment in which orientation-entanglement
has observable consequences.

The appropriate operational question in this context is: Whether rotating bodies in the physical space respect ${2\pi}$ periodicity
or ${4\pi}$? Consider, for example, a rock in an otherwise empty universe. If it is rotated by ${2\pi}$ radians about some axis,
then there is no reason to doubt that it will return back to its original state with no discernible effects \cite{Hartung}. This,
however, cannot be expected if there is at least one other object present in the universe \cite{Hartung}.
The rock will then have to rotate by
another ${2\pi}$ radians (i.e., a total of ${4\pi}$ radians) to return back to its original state relative to that other object,
as proved by the twist in the belt in Dirac's belt trick \cite{Penrose}. The twist shows that what is an identity transformation
for an isolated object is {\it not} an identity transformation for an object that is rotating relative to other
objects \cite{Hartung}. We can quantify this loss of identity by adapting a spinor representation of rotations.

Let the configuration space of all possible rotations of the rock be represented by the set ${S^3}$ of unit quaternions
(which, as we shall soon see, is a parallelized 3-sphere):
\begin{equation}
S^3:=\left\{\,{\bf q}(\psi,\,{\bf a}):=\exp{\left[\,{\boldsymbol\beta}({\bf a})\,\frac{\psi}{2}\,\right]}\;
\Bigg|\;||\,{\bf q}(\psi,\,{\bf a})\,||^2=1\right\}\!, \label{nonoonpara}
\end{equation}
where ${{\boldsymbol\beta}({\bf a})}$ is a bivector rotating about ${{\bf a}\in{\rm I\!R}^3}$
with the rotation angle ${\psi}$ in the range ${0\leq\psi < 4\pi}$. Throughout this paper we shall follow the notations,
conventions, and terminology of geometric algebra \cite{Clifford}\cite{Hestenes}. Accordingly,
${{\boldsymbol\beta}({\bf a})\in S^2\subset S^3}$ can be parameterized by a unit vector
${{\bf a}=a_1\,{\bf e}_1+a_2\,{\bf e}_2+a_3\,{\bf e}_3\in{\rm I\!R}^3}$ as
\begin{align}
{\boldsymbol\beta}({\bf a})\,&:=\,(\,I\cdot{\bf a}\,)\, \notag \\
&=\,a_1\,(\,I\cdot{\bf e}_1\,)
\,+\,a_2\,(\,I\cdot{\bf e}_2\,)\,+\,a_3\,(\,I\cdot{\bf e}_3\,) \notag \\
&=\,a_1\;{{\bf e}_2}\,\wedge\,{{\bf e}_3}
\,+\,a_2\;{{\bf e}_3}\,\wedge\,{{\bf e}_1}\,+\,a_3\;{{\bf e}_1}\,\wedge\,{{\bf e}_2}\,,
\end{align}
with ${{\boldsymbol\beta}^2({\bf a})=-1}$. Here the trivector ${I:={\bf e}_1\,\wedge\,{\bf e}_2\,\wedge\,{\bf e}_3}$
(which also squares to ${-1}$) represents a volume form of the physical space \cite{Clifford}.
Each configuration of the rock can thus be represented by a quaternion of the form
\begin{equation}
{\bf q}(\psi,\,{\bf a})\,=\,\cos\frac{\psi}{2}\,+\,{\boldsymbol\beta}({\bf a})\,\sin\frac{\psi}{2}\,, \label{defi-2}
\end{equation}
with ${\psi}$ being its rotation angle from ${{\bf q}(0,\,{\bf a})=1}$. More significantly for our purposes,
it is easy to check that ${{\bf q}(\psi,\,{\bf a})}$ respects the following rotational symmetries:
\begin{align}
{\bf q}(\psi+2\kappa\pi,\,{\bf a})\,&=\,-\,{\bf q}(\psi,\,{\bf a})\,\;\;\text{for}\;\,\kappa=1,3,5,7,\dots \label{clacct}\\
{\bf q}(\psi+4\kappa\pi,\,{\bf a})\,&=\,+\,{\bf q}(\psi,\,{\bf a})\,\;\;\text{for}\;\,\kappa=0,1,2,3,\dots \label{non-clacct}
\end{align}
Thus ${{\bf q}(\psi,\,{\bf a})}$ correctly represents the state of a rock that returns to itself only after even multiples
of a ${2\pi}$ rotation.
 
This is well and good as a mathematical representation of the orientation-entanglement \cite{Misner}, but can such changes
in sign have observable consequences? The answer to this question turns out to be in the affirmative \cite{Aharonov},
provided the ``rock''
happens to be microscopic and can be treated quantum mechanically \cite{Werner}. Despite the fact that physical quantities are
quadratic in the wave function, Aharonov and Susskind were able to demonstrate that in certain circumstances sign changes
of spinors under ${2\pi}$ rotations can lead to observable effects \cite{Aharonov}\cite{Werner}\cite{Weingard}. In particular,
they were able to show that if fermions are coherently shared between two spatially isolated systems, then a relative rotation
of ${2\pi}$ may be observable. They noted, however, that in classical physics relative as well as absolute ${2\kappa\pi}$ rotations
are unobservable \cite{Aharonov}. That is to say, in classical physics Eq.${\,}$(\ref{clacct}) has no observable
consequences \cite{Hartung}\cite{Weingard}.

It turns out, however, that this last conclusion is not quite correct. In what follows we shall demonstrate that in certain
circumstances Eq.${\,}$(\ref{clacct}) does lead to observable consequences in classical physics. This is not surprising when
one recalls the gimbal-lock singularity encountered in some representations of rotations, such as Euler angles. The singularity
arises because the group SO(3) of proper rotations in the physical space happens to be a connected but not a simply-connected
topological manifold \cite{Spinor}\cite{gimbal}.

To visualize this, imagine a solid ball of radius ${\pi}$. Let each point in this ball correspond to a rotation
with the direction from the origin of the ball representing the axis of rotation and the distance from
the origin representing the angle of rotation. Since rotations by ${+\,\pi}$ and ${-\,\pi}$
are the same, the antipodes of the surface of the ball must be identified with each other.
The ball therefore does not represent rotations in a continuous manner. Antipodal points of the ball are
considered equivalent even though they are far apart in the representation \cite{Spinor}.
This lack of continuity is the source of the gimbal-lock singularity.

This singularity can be eliminated \cite{gimbal}
by considering a second ball superimposed upon the first with coordinates that are a
mirror image of the first. The two surfaces are then glued together so that a rotation by ${+\,\pi}$
on the one ball is connected to the corresponding rotation by ${-\,\pi}$ on the other. This object is then
equivalent to ${S^3}$ in the same way that two discs glued together form ${S^2}$.
The topology is now represented by a simply-connected and continuous shape in four parameters, such as the set
\begin{equation}
\left\{\,\cos\frac{\psi}{2}\,,\;a_1\,\sin\frac{\psi}{2}\,,\;a_2\,\sin\frac{\psi}{2}\,,\;a_3\,\sin\frac{\psi}{2}\,\right\}
\label{roppara}
\end{equation}
appearing in definition (\ref{defi-2}). The gimbal-lock singularity is eliminated in this representation because each
rotation is now quantified, not only by axis and angle, but also by the relative orientation of the gimbal with respect to
its surroundings. Thus removal of a gimbal-lock provides a striking manifestation of orientation-entanglement in classical
physics. In what follows we shall demonstrate a similar macroscopic scenario, giving rise to
observable consequences of Eq.${\,}$(\ref{clacct}) in classical physics.

To this end, recall that the set of all unit quaternions satisfying Eqs.${\,}$(\ref{clacct}) and
(\ref{non-clacct}) forms the group SU(2), which is homeomorphic to the simply-connected space ${S^3}$ \cite{Analysis}.
This group is relevant in the macroscopic world when rotations of objects relative to other ``fixed'' objects are important.
On the other hand, purely local observations of rotations seem to be insensitive to the sign changes of the quaternions
constituting SU(2). In other words, the points ${-\,{\bf q}(\psi,\,{\bf a})}$ and ${+\,{\bf q}(\psi,\,{\bf a})}$ within
${\rm SU(2)}$ seem to represent one and the same rotation in the physical space ${{\rm I\!R}^3}$.
The group SO(3) is therefore obtained by identifying each quaternion ${-\,{\bf q}(\psi,\,{\bf a})}$ with its antipode
${+\,{\bf q}(\psi,\,{\bf a})}$ within SU(2)---{\it i.e.}, by identifying
the antipodal points of ${S^3}$. The space that results from this identification is the real projective space,
${{\rm I\!R}{\rm P}^3}$, which is a connected but not a simply-connected manifold. Consequently, the geodesic distances
${{\mathscr D}({\bf a},\,{\bf b})}$ between two quaternions ${{\bf q}(\psi_{\bf a},\,{\bf a})}$ and
${{\bf q}(\psi_{\bf b},\,{\bf b})}$ representing two different rotations within ${{\rm I\!R}^3}$ can be measurably different
on the manifolds ${{\rm SU(2)}\sim S^3}$ and ${{\rm SO(3)}\sim {\rm I\!R}{\rm P}^3}$. These distances would thus provide a
signature of spinorial sign changes between
${{\bf q}(\psi_{\bf a},\,{\bf a})}$ and ${{\bf q}(\psi_{\bf b},\,{\bf b})}$ within classical or macroscopic physics.

In the next two sections we therefore derive the relative geodesic distances on the manifolds SU(2) and SO(3). Then, in the
subsequent section, we sketch a macroscopic experiment, which, if realized, would allow to distinguish the distances on SU(2)
from those on SO(3) by observing correlations among a set of spin angular momenta. Then, in section V, we derive
the correlation function from the first principles, before concluding the paper in section VI.

\section{Geodesic Distance on ${\rm SU(2)}$}

In this section the geometry of a parallelized 3-sphere will play an important role.
To understand this geometry, consider the quadruple of real numbers defined in Eq.${\,}$(\ref{roppara}).
These numbers may be used to define a three-dimensional surface embedded in ${{\rm I\!R}^4}$,
homeomorphic to the 3-sphere of unit quaternions. This is possible because
there exists a bi-continuous one-to-one correspondence between the parameter space (\ref{roppara}) and the space
of unit quaternions. These two spaces are thus topologically equivalent \cite{Analysis}.

To see this, denote the points of the parameter space (\ref{roppara})
by the tips of the unit vectors ${{\bf Y}(\psi,\,{\bf a})\in{\rm I\!R}^4}$ as
\begin{align}
{\bf Y}(\psi,\,{\bf a})\,:=\,&\left(\,\cos\frac{\psi}{2}\,\right)\,{\bf e}_0\,+\,
\left(\,a_1\,\sin\frac{\psi}{2}\,\right)\,{\bf e}_1 \notag \\
&\,+\,\left(\,a_2\,\sin\frac{\psi}{2}\,\right)\,{\bf e}_2\,+\,\left(\,a_3\,\sin\frac{\psi}{2}\,\right)\,{\bf e}_3\,, \label{round}
\end{align}
where ${Y_0^2+Y_1^2+Y_2^2+Y_3^2=1}$ because ${a_1^2+a_2^2+a_3^2=1}$. ${{\bf Y}(\psi,\,{\bf a})}$ thus sweeps
the surface of a unit ball in ${{\rm I\!R}^4}$, and hence sweeps the points of a round 3-sphere of constant curvature and
vanishing torsion. It can be transformed into a spinor ${{\bf q}(\psi,\,{\bf a})\in{\rm I\!R}^4}$
representing a point of a ``flat'' 3-sphere as follows:
\begin{equation}
\left(\begin{array}{c} {\rm q}_0 \\ {\rm q}_1 \\ {\rm q}_2 \\ {\rm q}_3 \end{array}\right)\;=\;
\left(\begin{matrix} \; +\,{\bf e}_0 \; & \; 0 \; & \; 0 \; & \; 0 \; \\
                              \; 0 \; & \; +\,I \; & \; 0 \; & \; 0 \; \\
                               \; 0 \; & \; 0 \; & \; +\,I \; & \; 0 \; \\
                                \; 0 \; & \; 0 \; & \; 0 \; & \; +\,I \;
\end{matrix}\right)
\left(\begin{array}{c} {\rm Y}_0 \\ {\rm Y}_1 \\ {\rm Y}_2 \\ {\rm Y}_3 \end{array}\right)\!.
\end{equation}
This operation transforms the vector
\begin{equation}
{\bf q}(\psi,\,{\bf a}):={\rm q}_0\,{\bf e}_0+{\rm q}_1\,{\bf e}_1+{\rm q}_2\,{\bf e}_2+{\rm q}_3\,{\bf e}_3\in{\rm I\!R}^4
\end{equation}
into a unit quaternion:
\begin{align}
{\bf q}(\psi,\,{\bf a})\,&=\,\cos\frac{\psi}{2}\;\,{\bf e}_0\,{\bf e}_0\,+\,
a_1\,(\,I\cdot{\bf e}_1\,)\,\sin\frac{\psi}{2}\, \notag \\
&\;\;\;\;\;\;\;\;\;\;+\,a_2\,(\,I\cdot{\bf e}_2\,)\,\sin\frac{\psi}{2}\,+\,
a_3\,(\,I\cdot{\bf e}_3\,)\,\sin\frac{\psi}{2} \notag \\
&=\,\cos\frac{\psi}{2}\,+\,
a_1\,(\,{{\bf e}_2}\,\wedge\,{{\bf e}_3}\,)\,\sin\frac{\psi}{2}\, \notag \\
&\;\;\;\;\;+\,a_2\,(\,{{\bf e}_3}\,\wedge\,{{\bf e}_1}\,)\,\sin\frac{\psi}{2}\,+\,
a_3\,(\,{{\bf e}_1}\,\wedge\,{{\bf e}_2}\,)\sin\frac{\psi}{2} \notag \\
&=\,\cos\frac{\psi}{2}\,+\,{\boldsymbol\beta}({\bf a})\,\sin\frac{\psi}{2} \notag \\
&=\,\exp{\left[\,{\boldsymbol\beta}({\bf a})\,\frac{\psi}{2}\,\right]}\,,
\label{flat}
\end{align}
where ${{\bf q}(\psi,\,{\bf a})\,{\bf q}^{\dagger}(\psi,\,{\bf a})=||\,{\bf q}(\psi,\,{\bf a})\,||^2=1}$ since we have
${{\bf e}_0\,{\bf e}_0={\bf e}_0\cdot{\bf e}_0+{\bf e}_0\wedge{\bf e}_0=1}$
and ${{\boldsymbol\beta}({\bf a}){\boldsymbol\beta}^{\dagger}({\bf a})=1}$.
As we shall see, ${{\bf q}(\psi,\,{\bf a})}$
represents a point of a 3-sphere that is ``flat'', but exhibits non-zero and constant torsion.

The reciprocal relationship between the 4-vector (\ref{round}) and the spinor (\ref{flat}) can now be succinctly described${\;}$as
\begin{equation}
{\rm Y}_i\,=\,\gamma^{\dagger}_{ij}\,{\rm q}_j \;\;\;\;\;\;\text{and}\;\;\;\;\;\; {\rm q}_i\,=\,\gamma_{ij}\,{\rm Y}_j \,,
\label{sev}
\end{equation}
where
\begin{equation}
\gamma_{ij}^{\dagger}\;=\,
\left(\begin{matrix} \; +\,{\bf e}_0 \; & \; 0 \; & \; 0 \; & \; 0 \; \\
                     \; 0 \; & \; -\,I \; & \; 0 \; & \; 0 \; \\                                                                  
                     \; 0 \; & \; 0 \; & \; -\,I \; & \; 0 \; \\                                                                  
                     \; 0 \; & \; 0 \; & \; 0 \; & \; -\,I \;
\end{matrix}\right)
\end{equation}
and
\begin{equation}
\gamma_{ij}\;=\,
\left(\begin{matrix} \; +\,{\bf e}_0 \; & \; 0 \; & \; 0 \; & \; 0 \; \\
                     \; 0 \; & \; +\,I \; & \; 0 \; & \; 0 \; \\                                                                  
                     \; 0 \; & \; 0 \; & \; +\,I \; & \; 0 \; \\                                                                  
                     \; 0 \; & \; 0 \; & \; 0 \; & \; +\,I \;
\end{matrix}\right)
\end{equation}
with ${\gamma_{ij}^{\dagger}\,\gamma_{ij}\,=\,{\rm id}}$. The transformation-maps
${{\rm Y}_i=\gamma^{\dagger}_{ij}\,{\rm q}_j}$ and ${{\rm q}_i=\gamma_{ij}\,{\rm Y}_j}$ are thus smooth bijections
${{\bf Y}: S^3_F\rightarrow S^3_R}$
and ${{\bf q}: S^3_R\rightarrow S^3_F}$ mentioned above, with ${S^3_F}$ representing
the flat 3-sphere and ${S^3_R}$ representing the round 3-sphere.  

It is important to note here that the transformation of the 4-vectors ${{\bf Y}(\psi,\,{\bf a})}$ defined in (\ref{round})
into the spinors ${{\bf q}(\psi,\,{\bf a})}$ defined in (\ref{flat}) induces dramatic differences in the geometry and topology
of the 3-sphere. The round 3-sphere, charted by ${{\bf Y}(\psi,\,{\bf a})}$, is known to have constant curvature
but vanishing torsion. This is the
3-sphere that appears in the Friedmann-Robertson-Walker solution of Einstein's field equations \cite{Walker}.
The flat 3-sphere, charted by ${{\bf q}(\psi,\,{\bf a})}$, on the other hand, has vanishing curvature but constant torsion.
This is the 3-sphere that appears in the teleparallel gravity \cite{telepar}.
In other words, the 3-sphere constituted by quaternions
is parallelized---by the very algebra of quaternions---to be {\it absolutely} flat \cite{Eisenhart}\cite{Nakahara}.

\begin{figure}
\hrule
\scalebox{1}{
\begin{pspicture}(0.0,-4.2)(4.9,3.7)

\pscircle[linewidth=0.4mm,linecolor=blue](2.5,-0.45){3.0}

\psline[linewidth=0.2mm,linestyle=dashed](-0.4,-0.45)(5.4,-0.45)

\psellipticarc[linewidth=0.3mm,linestyle=dashed,linecolor=blue](2.5,-0.45)(2.96,1.35){3}{49.1}

\psellipticarc[linewidth=0.3mm,linestyle=dashed,linecolor=blue](2.5,-0.45)(2.96,1.35){90.5}{361}

\psellipticarc[linewidth=0.4mm,linecolor=green,arrowinset=0.3,arrowsize=2pt 3,arrowlength=1]{->}(2.5,-0.45)(2.95,1.35){198.5}{256}

\psellipticarc[linewidth=0.3mm,arrowinset=0.3,arrowsize=2pt 3,arrowlength=1]{<->}(2.5,-0.45)(1.1,0.5){198}{256}

\psline[linewidth=0.3mm,arrowinset=0.3,arrowsize=3pt 3,arrowlength=2]{->}(2.5,-0.45)(2.17,-1.77)

\psline[linewidth=0.3mm,arrowinset=0.3,arrowsize=3pt 3,arrowlength=2]{->}(2.5,-0.45)(0.082,-1.22)

\psline[linewidth=0.4mm,linecolor=red,arrowinset=0.3,arrowsize=2pt 3,arrowlength=1]{->}(0.082,-1.22)(0.7,-1.62)

\psellipticarc[linewidth=0.4mm,linecolor=green,arrowinset=0.3,arrowsize=2pt 3,arrowlength=1]{<-}(2.5,-0.45)(1.8,2.2){50.8}{90}

\psellipticarc[linewidth=0.3mm,arrowinset=0.3,arrowsize=2pt 3,arrowlength=1]{<->}(2.5,-0.45)(0.8,0.9){50.3}{90}

\psline[linewidth=0.3mm,arrowinset=0.3,arrowsize=3pt 3,arrowlength=2]{->}(2.5,-0.45)(2.5,1.75)

\psline[linewidth=0.4mm,linecolor=red,arrowinset=0.3,arrowsize=2pt 3,arrowlength=1]{->}(2.5,1.76)(3.3,1.66)

\psline[linewidth=0.3mm,arrowinset=0.3,arrowsize=3pt 3,arrowlength=2]{->}(2.5,-0.45)(3.75,1.1)

\rput{330}(3.95,0.25){{${\bf q(a'\,)}$}}

\rput{350}(1.9,1.25){{${\bf q(a)}$}}

\rput{350}(2.85,0.65){{${d\chi}$}}

\rput{338}(3.35,1.85){{${d{\bf q\!={\bf t}_{\bf q}}\,d\chi}$}}

\rput{343}(0.91,-1.88){{${d(I\cdot{\bf a})={\bf t}_{\bf a}\,d\eta}$}}

\rput{343}(1.75,-1.12){{${d\eta}$}}

\rput{357}(0.6,-0.7){{${I\cdot{\bf a}}$}}

\rput{357}(2.8,-1.2){{${I\cdot{\bf a'}}$}}

\psdot*(2.5,-0.45)

\put(-1.27,1.55){{\large ${S^3}$}}

\psarc[linewidth=0.3mm,arrowinset=0.3,arrowsize=2pt 3,arrowlength=1]{<-}(-0.7,0.95){0.7}{30}{90}

\put(5.7,-2.1){{\large ${S^2}$}}

\psarc[linewidth=0.3mm,arrowinset=0.3,arrowsize=2pt 3,arrowlength=1]{<-}(5.4,-1.25){0.7}{195}{240}

\psarc[linewidth=0.3mm](5.4,-1.25){0.7}{251.5}{285}

\end{pspicture}}
\vspace{0.25cm}
\hrule
\caption{The relation between the angle ${\chi}$ between ${\bf q(a)}$ and
${\bf q(a'\,)\,}$ and ${\eta}$ between ${\,I\cdot{\bf a}\,}$ and ${\,I\cdot{\bf a'}}$ is non-linear in general.}
\label{fig-798}
\vspace{0.3cm}
\hrule
\end{figure}
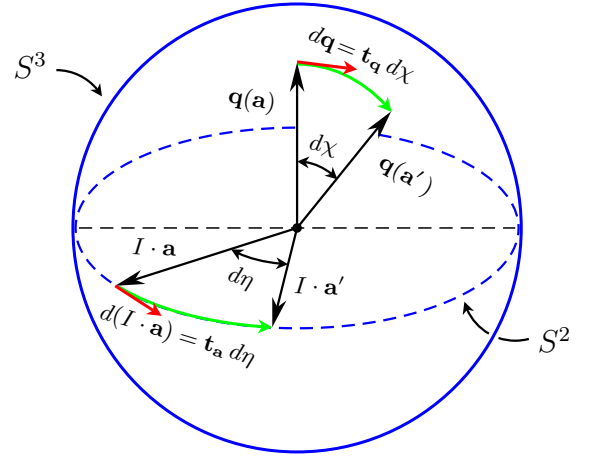

To understand the topological difference between these two 3-spheres, let us bring out the
round metric hidden in Eq.${\,}$(\ref{round})
by expressing the components of the vector ${\bf a}$ in the polar coordinates of an equatorial 2-sphere:
\begin{align}
a_1\,&=\,\sin\theta\,\cos\phi\,, \\
a_2\,&=\,\sin\theta\,\sin\phi\,, \\
a_3\,&=\,\cos\theta\,.
\end{align}
In terms of these components the components of ${{\bf Y}(\psi,\,{\bf a})}$ take the form
\begin{align}
Y_0\,&=\,\cos\chi\,, \\
Y_1\,&=\,\sin\chi\,\sin\theta\,\cos\phi\,, \\
Y_2\,&=\,\sin\chi\,\sin\theta\,\sin\phi\,, \\
Y_3\,&=\,\sin\chi\,\cos\theta\,.
\end{align}
where we have set ${\psi/2=\chi}$ for convenience. Note that ${\theta}$ ranges from ${0}$ to ${\pi}$, whereas ${\chi}$ and ${\phi}$
range from ${0}$ to ${2\pi}$. The corresponding line element can now be calculated from the differentials
\begin{align}
dY_0\,&=\,-\,\sin\chi\,d\chi\,, \\
dY_1\,&=\,\cos\chi\,\sin\theta\,\cos\phi\;d\chi\,+\,\sin\chi\,\cos\theta\,\cos\phi\;d\theta\, \notag \\
&\;\;\;\;\;\;\;\;\;\;\;\;\;\;\;\;\;\;\;\;\;\;\;\;-\,\sin\chi\,\sin\theta\,\sin\phi\;d\phi\,,\\
dY_2\,&=\,\cos\chi\,\sin\theta\,\sin\phi\;d\chi\,+\,\sin\chi\,\cos\theta\,\sin\phi\;d\theta\, \notag \\
&\;\;\;\;\;\;\;\;\;\;\;\;\;\;\;\;\;\;\;\;\;\;\;\;+\,\sin\chi\,\sin\theta\,\cos\phi\;d\phi\,, \\
dY_3\,&=\,\cos\chi\,\cos\theta\;d\chi\,-\,\sin\chi\,\sin\theta\;d\theta\,.
\end{align}
Moreover, from the normalization condition
\begin{equation}
Y_0^2\,+\,Y_1^2\,+\,Y_2^2\,+\,Y_3^2\,=\,1^2\,=\,1
\end{equation}
we also have
\begin{equation}
2Y_0\,dY_0\,+\,2Y_1\,dY_1\,+\,2Y_2\,dY_2\,+\,2Y_3\,dY_3\,=\,0\,,
\end{equation}
which allows us to express ${dY_0}$ in terms of ${dY_1}$, ${dY_2}$, and ${dY_3}$. Then, in the
hyper-spherical coordinates ${(\,\chi,\,\theta,\,\phi\,)}$, the line element on the 3-sphere works out to be
\begin{align}
ds^2\,&=\,g(d{\bf Y},\,d{\bf Y}) \notag \\
&=\,dY^2_0\,+\,dY^2_1\,+\,dY^2_2\,+\,dY^2_3 \notag \\
&=\,d\chi^2\,+\,\sin^2\chi\left[\,d\theta^2+\,\sin^2\theta\,d\phi^2\,\right]. \label{FRW-metric}
\end{align}
This is the Friedmann-Robertson-Walker line element \cite{Misner} representing a 3-sphere embedded in a
four-dimensional Euclidean space, with constant curvature and vanishing torsion:
\begin{equation}
{\mathscr R}^{\,\alpha}_{\;\;\,\beta\,\gamma\,\delta}\,\not=\,0\;\;\;\;\;\;\;\;\text{but}
\;\;\;\;\;\;\;\;{\mathscr T}_{\,\alpha\,\beta}^{\,\gamma}\,=\,0\,.\label{vasin}
\end{equation}
This metric, however, does not provide a single-valued coordinate chart over the entire 3-sphere \cite{Ryder}.
In going from the north pole (${\chi=0}$) to the equator (${\chi=\pi/2}$), the variable ${\sin\chi}$
ranges from 0 to 1; however, in going from the equator to the south pole of the sphere (${\chi=\pi}$), it runs backwards from 1 to
0. Thus the space ${S^3_R}$ is charted by the coordinates that are not single-valued in ${\sin\chi}$. This can be seen
more clearly by setting ${\sin\chi=r}$ and rewriting the line element (\ref{FRW-metric}) as
\begin{equation}
ds^2\,=\,\frac{\,dr^2}{1-r^2}\,+\,r^2\left[\,d\theta^2+\,\sin^2\theta\,d\phi^2\,\right]. \label{non-metric}
\end{equation}
It is now easy to appreciated that there is a singularity in this metric at ${r=1}$. This is of course
the well known coordinate singularity which cannot be eliminated by a mere change of variables.
It {\it can}${\,}$ be eliminated, however, by parallelizing or ``flattening'' the 3-sphere with respect to
a set of quaternionic bases \cite{Nakahara}. 

To understand this, let ${T_{\bf q}S^3}$ denote the tangent space to ${S^3}$ at the tip of the unit
spinor ${{\bf q} \in {\rm I\!R}^4}$, defined by
\begin{equation}
T_{\bf q}S^3:=\left\{({\bf q},\,{\bf t_q})\;\Big|\;
{\bf q},\,{\bf t_q}\in {\rm I\!R}^4,\,||{\bf q}||=1,\,{\bf q}\cdot {\bf t}_{\bf q}^{\dagger}=\,0\right\}\!,
\label{tanspace}
\end{equation}
where ${{\bf q}\cdot {\bf t}_{\bf q}^{\dagger}}$ represents the inner product between ${\bf q}$ and ${{\bf t}_{\bf q}^{\dagger}}$
with ${{\bf t_q}:=d{\bf q}/d\psi}$.
Then, denoting the tip of ${\bf q}$ by ${q={\bf q}\cap S^3}$, the tangent bundle of ${S^3}$ can be expressed as
\begin{equation}
{\rm T}S^3\,:=\!\bigcup_{\,q\,\in\, S^3}\{\,q\,\}\times T_{\bf q}S^3.\label{tanbundle}
\end{equation}
Now this tangent bundle happens to be ${trivial\,}$:
\begin{equation}
{\rm T}S^3\,\equiv\,S^3\times{\rm I\!R}^3.\label{cfeq3}
\end{equation}
The triviality of the bundle ${{\rm T}S^3}$ means that the 3-sphere is {\it parallelizable} \cite{Eisenhart}\cite{Nakahara}.
A ${d}$-dimensional manifold
is said to be parallelizable if it admits ${d}$ vector fields that are linearly-independent everywhere. On a 3-sphere we can
always find three linearly-independent vector fields that are nowhere vanishing. These vector fields can then be used to define
a basis of the tangent space at each of its points. Thus, a global anholonomic frame can be defined on the 3-sphere that
fixes each of its points uniquely.

The parallelizability of the 3-sphere, however, is not guaranteed for all of its representations, since there are more than one
ways to embed one space into another. For example, one may consider embedding ${S^3}$ into ${{\rm I\!R}^4}$ by means of the vector
field ${{\bf Y}(\,\chi,\,\theta,\,\phi\,)}$
defined in Eq.${\,}$(\ref{round}), but as we saw above the resulting representation, namely ${S^3_R}$, would not be a parallelized
sphere \cite{Nakahara}. Given three linearly-independent vector fields forming a basis of the tangent space at one point of
${S^3_R\,}$, say at ${(\,\chi,\,\theta,\,\phi\,)}$, it would not be possible to find three linearly-independent
vector fields forming a basis of the tangent space at every other point of ${S^3_R}$. This turns out to be
possible, however, if we switch from the vector field ${{\bf Y}(\,\chi,\,\theta,\,\phi\,)}$
to the spinor field ${{\bf q}(\,\chi,\,\theta,\,\phi\,)}$ defined by Eq.${\,}$(\ref{flat})
with the property (\ref{clacct}).

Suppose we are given a tangent space at the tip of a spinor
${\,{\bf q}_0=(1,\,0,\,0,\,0)\in{\rm I\!R}^4\,}$, spanned by the basis
\begin{align}
&\left\{\;\beta_1({\bf q}_0),\;\beta_2({\bf q}_0),\;\beta_3({\bf q}_0)\,\right\}\, \notag \\
&\;\;\;\;\;\;\;\;\;\;\;\;\;\;\;\;\;\;\;\;\equiv\,
\left\{\,{{\bf e}_2}\,\wedge\,{{\bf e}_3},\,\;{{\bf e}_3}\,\wedge\,{{\bf e}_1},\,\;{{\bf e}_1}\,\wedge\,{{\bf e}_2}\,\right\},
\label{babablack}
\end{align}
with the base bivectors 
\begin{align}
\beta_1({\bf q}_0)&=(\,0,\,\;1,\,\;0,\,\;0), \label{inthemeri-1} \\
\beta_2({\bf q}_0)&=(\,0,\,\;0,\,\;1,\,\;0), \label{inthemeri-2} \\
\text{and}\;\;\;\;\beta_3({\bf q}_0)&=(\,0,\,\;0,\,\;0,\,\;1) \label{inthemeri-3}
\end{align}
satisfying the inner product
\begin{align}
\langle\beta_{\mu}({\bf q}_0),\,\beta^{\dagger}_{\nu}({\bf q}_0)\rangle
\,=\,\beta_{\mu}({\bf q}_0)\cdot\beta^{\dagger}_{\nu}({\bf q}_0)\,=\,\delta_{\mu\nu}\,,\label{orige-flat}
\end{align}
where ${\mu,\,\nu=1,\,2,\,3}$ and ${\langle\beta_{\mu},\,\beta^{\dagger}_{\nu}\rangle}$ is defined by the map
\begin{equation}
\langle\,\cdot\;,\,\cdot\,\rangle: T_pS^3\times T^{*}_pS^3 \rightarrow {\rm I\!R}\,,
\end{equation}
with ${p={\bf p}\cap S^3}$ (here ${T^{*}_pS^3}$ is the cotangent space at ${p}$ \cite{Nakahara}).
This basis would allow us to express any arbitrary tangent bivector at the tip of ${{\bf q}_0}$ as
\begin{equation}
I\cdot{\bf n}\,=\,n_1\;{{\bf e}_2}\,\wedge\,{{\bf e}_3}
\,+\,n_2\;{{\bf e}_3}\,\wedge\,{{\bf e}_1}
\,+\,n_3\;{{\bf e}_1}\,\wedge\,{{\bf e}_2}\,.
\end{equation}
Then the tangent bases ${\left\{\,\beta_1({\bf q}),\;\beta_2({\bf q}),\;\beta_3({\bf q})\right\}}$ at the tip of
any ${{\bf q}\in S^3}$
can be found by taking a geometric product of the basis (\ref{babablack}) with ${\bf q}$ using the bivector subalgebra
\begin{equation}
(I\cdot{\bf e}_{\mu})\,(I\cdot{\bf e}_{\nu})\,=\,-\;\delta_{{\mu}{\nu}}
\,-\,\epsilon_{{\mu}{\nu}{\rho}}\,(I\cdot{\bf e}_{\rho})\,\label{66666}\end{equation}
(with repeated Greek indices summed over), which gives
\begin{align}
&\beta_1({\bf q})=({{\bf e}_2}\,\wedge\,{{\bf e}_3})\,{\bf q} \notag \\
&\;\;=\,-\,q_1\,+\,q_0\,({{\bf e}_2}\,\wedge\,{{\bf e}_3})\,+\,q_3\,({{\bf e}_3}\,\wedge\,{{\bf e}_1})
\,-\,q_2\,({{\bf e}_1}\,\wedge\,{{\bf e}_2}) \notag \\
&\;\;=(-\,q_1,\,\;q_0,\,\;q_3,\,\;-\,q_2), \label{inthederi-1} \\
&\beta_2({\bf q})=({{\bf e}_3}\,\wedge\,{{\bf e}_1})\,{\bf q} \notag \\
&\;\;=\,-\,q_2\,-\,q_3\,({{\bf e}_2}\,\wedge\,{{\bf e}_3})\,+\,q_0\,({{\bf e}_3}\,\wedge\,{{\bf e}_1})
\,+\,q_1\,({{\bf e}_1}\,\wedge\,{{\bf e}_2}) \notag \\
&\;\;=(-\,q_2,\,\;-\,q_3,\,\;q_0,\,\;q_1), \label{inthederi-2} \\
&\beta_3({\bf q})=({{\bf e}_1}\,\wedge\,{{\bf e}_2})\,{\bf q} \notag \\
&\;\;=\,-\,q_3\,+\,q_2\,({{\bf e}_2}\,\wedge\,{{\bf e}_3})\,-\,q_1\,({{\bf e}_3}\,\wedge\,{{\bf e}_1})
\,+\,q_0\,({{\bf e}_1}\,\wedge\,{{\bf e}_2}) \notag \\
&\;\;=(-\,q_3,\,\;q_2,\,\;-\,q_1,\,\;q_0). \label{inthederi-3}
\end{align}
It is easy to check that the bases ${\left\{\,\beta_1({\bf q}),\;\beta_2({\bf q}),\;\beta_3({\bf q})\right\}}$
are indeed orthonormal for all ${\bf q}$ with respect to the usual inner product in ${{\rm I\!R}^4}$, with each of the three
basis elements ${\beta_{\mu}({\bf q})}$ also being
orthogonal to ${{\bf q}=(q_0,\,\;q_1,\,\;q_2,\,\;q_3)}$, and thus define a tangent space ${{\rm I\!R}^3}$ at the tip of that
${\bf q}$. This procedure of finding
orthonormal tangent bases at any point of ${S^3 }$ can be repeated {\it ad infinitum}, providing a continuous
field of absolutely parallel spinorial tangent vectors at every point of ${S^3}$. That is, given the basis
${\left\{\,\beta_1({\bf q}),\;\beta_2({\bf q}),\;\beta_3({\bf q})\right\}}$ of the tangent space ${T_{\bf q}S^3}$ at
the tip of some spinor ${{\bf q}\in{\rm I\!R}^4}$, the basis of the tangent space ${T_{\bf p}S^3}$ at the tip of any
other spinor ${{\bf p}\in{\rm I\!R}^4}$ can be found by computing 
\begin{equation}
\left\{\;\beta_1({\bf p}),\;\beta_2({\bf p}),\;\beta_3({\bf p})\,\right\}=
\left\{\;\beta_1({\bf q})\,{\bf p},\;\beta_2({\bf q})\,{\bf p},\;\beta_3({\bf q})\,{\bf p}\,\right\}, \label{inthemeri}
\end{equation}
and so on for {\it all} ${\,}$points of ${S^3}$. This allows us to generate a continuous, orthonormality preserving
translation of the tangent basis at ${\bf q}$ to tangent basis at ${\bf p}$, for {\it all} ${\,}$pairs of spinors ${\bf q}$
and ${\bf p}$. In other words, for every ${{\bf p}\in S^3}$ the set
${\left\{\;\beta_1({\bf p}),\;\beta_2({\bf p}),\;\beta_3({\bf p})\,\right\}}$ is a basis of ${T_{\bf p}S^3}$.
As a result, each point ${{\bf p}\in S^3}$ is characterized by a tangent spinor ${\bf q}$ of the form (\ref{flat}),
representing a smooth flowing motion of that point, without any discontinuities, singularities, or fixed points hindering
its coordinatization.

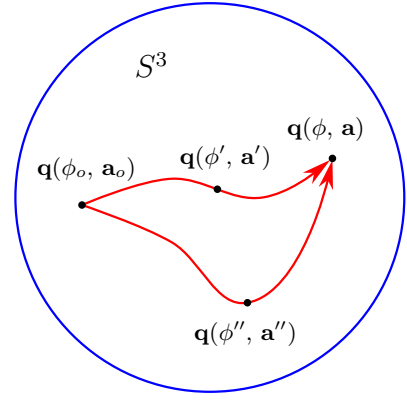
\begin{figure}
\hrule
\scalebox{1}{
\begin{pspicture}(-5.9,-3.7)(2.5,2.8)

\pscurve[linewidth=0.3mm,linecolor=red,arrowinset=0.3,arrowsize=3pt 3,arrowlength=2]{->}(-3.5,-0.55)(-2.3,-0.2)(-1.1,-0.45)(-0.2,0.05)

\pscurve[linewidth=0.3mm,linecolor=red,arrowinset=0.3,arrowsize=3pt 3,arrowlength=2]{->}(-3.5,-0.55)(-2.3,-1.05)(-1.3,-1.85)(-0.17,0.05)

\pscircle[linewidth=0.3mm,linecolor=blue](-1.8,-0.45){2.6}

\put(-2.8,1.15){{\large ${S^3}$}}

\put(-2.2,0.0){{${{\bf q}(\phi',\,{\bf a'})}$}}

\put(-2.01,-2.35){{${{\bf q}(\phi'',\,{\bf a''})}$}}

\put(-4.1,-0.15){{${{\bf q}(\phi_o,\,{\bf a}_o)}$}}

\put(-0.77,0.38){{${{\bf q}(\phi,\,{\bf a})}$}}

\pscircle*(-1.7,-0.34){0.05}

\pscircle*(-1.3,-1.85){0.05}

\pscircle*(-3.5,-0.55){0.05}

\pscircle*(-0.17,0.07){0.05}

\end{pspicture}}
\hrule
\caption{Path-independence of parallel transports within ${S^3}$.}
\vspace{0.3cm}
\label{fig-512}
\hrule
\end{figure}

One immediate consequence of the above construction of orthonormal basis for the
tangent space at each point  ${{\bf q}\in S^3}$ is that it renders the metric tensor on ${S^3}$ ``flat'',
\begin{equation}
g(\,\beta_{\mu}({\bf q}){\hspace{1pt}},\;\beta^{\dagger}_{\nu}({\bf q}))\,=\,\beta_{\mu}({\bf q})\cdot\beta^{\dagger}_{\nu}({\bf q})
\,=\,\delta_{\mu\nu}\,,\label{flat-me}
\end{equation}
extending the inner product (\ref{orige-flat}) of the tangent basis at ${{\bf q}_0}$ to all points ${\bf q}$ of ${S^3}$.
Consequently (and contrary to the misleading impressions given by the Figs.${\,}$\ref{fig-798} and \ref{fig-512}),
the Riemann curvature tensor of ${S^3}$ vanishes identically,
\begin{equation}
{\mathscr R}^{\,\alpha}_{\;\;\,\beta\,\gamma\,\delta}\,=\,\partial_\gamma\,\Omega_{\delta\,\beta}^{\alpha}\,-\,
\partial_\delta\,\Omega_{\gamma\,\beta}^{\alpha}\,+\,\Omega_{\delta\,\beta}^{\rho}\,\Omega_{\gamma\,\rho}^{\alpha}\,-\,
\Omega_{\gamma\,\beta}^{\rho}\,\Omega_{\delta\,\rho}^{\alpha}\,=\,0\,,
\end{equation}
with respect to the a-symmetric Weitzenb\"ock connection
\begin{equation}
\Omega_{\alpha\,\beta}^{\gamma}\,=\,\Gamma_{\alpha\,\beta}^{\,\gamma}\,+\,
{\mathscr T}_{\,\alpha\,\beta}^{\,\gamma}\,, \label{W-con}
\end{equation}
where ${\Gamma_{\alpha\,\beta}^{\,\gamma}}$ is the symmetric Levi-Civita connection and
${{\mathscr T}_{\,\alpha\,\beta}^{\,\gamma}}$
is the totally anti-symmetric torsion tensor. This vanishing of the curvature tensor renders the resulting parallelism on ${S^3}$
absolute---{\it i.e.}, it guarantees the {\it path-independence}${\,}$ of the parallel transport within ${S^3}$. As a result, a
parallel transport of arbitrary spinor (or tensor) within ${S^3}$ is simply a translation of that spinor within ${S^3}$.
In particular, the path ${(\phi_o,\,{\bf a}_o)\,\rightarrow\,(\phi',\,{\bf a'})\,\rightarrow\,(\phi,\,{\bf a})}$
shown in Fig.${\,}$\ref{fig-512} is a path along which the quaternion ${{\bf q}(\phi_o,\,{\bf a}_o)}$ is translated, say,
from ${{\bf q}(\phi_o,\,{\bf a}_o)\in T_{(\phi_o,\,{\bf a}_o)}S^3}$ to
${{\bf q}(\phi,\,{\bf a})\in T_{(\phi,\,{\bf a})}S^3}$. For each path chosen for such a translation, say through the
point ${(\phi'',\,{\bf a''})}$ instead of the point ${(\phi',\,{\bf a'})}$,
the end result would be the same, namely the quaternion ${{\bf q}(\phi,\,{\bf a})}$ tangential to the point
${(\phi,\,{\bf a})\in S^3}$. In other
words, the end result would be {\it path-independent}, thanks to the vanishing of
${{\mathscr R}^{\,\alpha}_{\;\;\,\beta\,\gamma\,\delta}\,}$.

However, since ${S^3}$ is not a Euclidean space \cite{Eberlein},
for the above path-independence to be possible the torsion within ${S^3}$ must
necessarily be non-vanishing. In fact it is straightforward to verify that the torsion within ${S^3_F}$ is indeed
non-zero and constant. For two arbitrary spinors ${I\cdot{\bf a}=a_{\mu}{\beta}_{\mu}}$ and
${I\cdot{\bf b}=b_{\nu}{\beta}_{\nu}}$ belonging to the tangent space ${T_{\bf p}S^3}$,
the torsion tensor ${\mathscr T}$ can be written as
\begin{equation}
{\mathscr T}(I\cdot{\bf a},\,I\cdot{\bf b})\,=\,a^{\mu}b^{\nu}\left\{\,\nabla_{\!\!{\beta}_{\mu}}{\beta}_{\nu}
\,-\,\nabla_{\!\!{\beta}_{\nu}}{\beta}_{\mu}\,-\,\left[\,{\beta}_{\mu},\,{\beta}_{\nu}\right]\,\right\},
\end{equation}
where the Lie bracket ${[\,\cdot\;,\,\cdot\,]}$ in the last term is the same as the commutator product bracket for bivectors. 
But since ${\left\{\,\beta_1({\bf p}),\;\beta_2({\bf p}),\;\beta_3({\bf p})\right\}}$ defines a ``frame field'' at the tip
of every ${{\bf p}\in{\rm I\!R}^4}$, each base bivector ${{\beta}_{\mu}}$ remains constant upon parallel
transport relative to this frame field, giving
${\nabla_{\!\!{\beta}_{\mu}}{\beta}_{\nu}=0=\nabla_{\!\!{\beta}_{\nu}}{\beta}_{\mu}}$.
Consequently, the above expression of the torsion tensor simplifies to
\begin{align}
{\mathscr T}(I\cdot{\bf a},\,I\cdot{\bf b})\,&=\,-\,a_{\mu}b_{\nu}\left[\,{\beta}_{\mu},\,{\beta}_{\nu}\right] \notag \\
\,&=\,-\,\left[\,(\,I\cdot{\bf a}\,),\,(\,I\cdot{\bf b}\,)\,\right] \notag \\
\,&=\,I\cdot({\bf a}\times{\bf b})\,=\,{\bf a}\,\wedge\,{\bf b}\,. \label{torsioninthe}
\end{align}
In other words, instead of (\ref{vasin}), we now have vanishing curvature but non-vanishing torsion \cite{Christian}:
\begin{equation}
{\mathscr R}^{\,\alpha}_{\;\;\,\beta\,\gamma\,\delta}\,=\,0\;\;\;\;\;\;\;\;\text{but}
\;\;\;\;\;\;\;\;{\mathscr T}_{\,\alpha\,\beta}^{\,\gamma}\,\not=\,0\,.
\end{equation}
This completes the transformation of the round 3-sphere ${S^3_R}$ of Eq.${\,}$(\ref{FRW-metric}) into the flat 3-sphere
${S^3_F}$ of Eq.${\,}$(\ref{flat-me}). 

As noted above, for us the importance of parallelizing ${S^3}$ and the corresponding vanishing of its Riemann tensor
lies in the availability of the orthonormality preserving continuous transport of the basis of ${T_{\bf p}S^3}$ to
${T_{\bf q}S^3}$ in a path-independent manner. This enables us to introduce the notion of parallelity of vectors tangent
to ${S^3}$ at {\it any} two points ${p,\,q}$ in ${S^3}$ in an absolute manner, thus allowing unambiguous distant comparison
between the directions of tangent spinors at {\it different} points of ${S^3}$. Consequently, if
${\{a_{\mu}\;{\beta}_{\mu}({\bf p})\}}$ and ${\{b_{\nu}\;{\beta}_{\nu}({\bf q})\}}$ are two
bivectors belonging to two different tangent spaces ${T_{\bf p}S^3}$ and ${T_{\bf q}S^3}$ at two different
points of ${S^3}$, then their parallelity allows us to compute the inner product between them as follows:
\begin{align}
\langle\,\{a_{\mu}\;{\beta}_{\mu}({\bf p})\},\;\{b_{\nu}\;{\beta}_{\nu}({\bf q})\}\,\rangle\,&=\,
\langle\,\{a_{\mu}\;{\beta}_{\mu}({\bf q})\},\;\{b_{\nu}\;{\beta}_{\nu}({\bf q})\}\,\rangle \notag \\
&=\,-\,\beta_{\mu}({\bf q})\cdot\beta^{\dagger}_{\nu}({\bf q})\;a_{\mu}\,b_{\nu}\, \notag \\
&=\,-\,\delta_{\mu\nu}\,a_{\mu}\,b_{\nu}\, \notag \\
&=\,-\,\cos\eta_{{\bf a}{\bf b}}\, \notag \\
&=\,-\,\cos\frac{\psi_{{\bf a}{\bf b}}}{2}\,, \label{conpat}
\end{align}
where ${\psi_{{\bf a}{\bf b}}}$ is the amount of rotation required to align the bivector
${\{\,a_{\mu}\,{\beta}_{\mu}({\bf q})\,\}}$ with the bivector ${\{\,b_{\nu}\,{\beta}_{\nu}({\bf q})\,\}}$.
Thus the inner product on ${S^3}$ itself provides a unique distance measure between two rotations in the physical space
with ${4\pi}$ periodicity ({\it i.e.,} spinorial sensitivity). More precisely,
\begin{align}
{\mathscr D}({\bf a},\,{\bf b})\,&=\,-\,\cos\eta_{{\bf a}{\bf b}} \notag \\
&=\,\text{a geodesic distance on ${S^3\!\sim{\rm SU(2)}}$}\,, \label{dismes}
\end{align}
where ${\eta_{{\bf a}{\bf b}}}$ is half of the rotation angle described above.

Although evident from both parameterizations (\ref{defi-2}) and (\ref{round}),
this distance measure may seem rather unusual. It may seem to be a consequence of the parallelization of ${S^3}$ rather than
a natural choice. In fact it is both physically and mathematically compelling \cite{Hestenes}, not the least because
the geodesic equation for a round ${S^3}$ with ${{\mathscr R}^{\,\alpha}_{\;\;\,\beta\,\gamma\,\delta}\not=0}$ and
${{\mathscr T}_{\,\alpha\,\beta}^{\,\gamma}=0}$ yields the {\it same} physical trajectory as the geodesic equation
for a parallelized ${S^3}$ with ${{\mathscr R}^{\,\alpha}_{\;\;\,\beta\,\gamma\,\delta}=0}$ and
${{\mathscr T}_{\,\alpha\,\beta}^{\,\gamma}\not=0}$. Moreover, in the next section we shall see that
${-\cos\eta_{{\bf a}{\bf b}}}$ specifies the correct geodesic distance on SU(2), because it is
simply a horizontal lift of the geodesic distance associated with a natural metric on SO(3)---{\it i.e.}, 
the configuration space of all proper rotations in ${{\rm I\!R}^3}$.

To appreciate this, note that in a manifold with torsion geodesics are not the straight lines. In the presence of torsion
the geodesic equation differs from the autoparallel equation and takes the form
\begin{equation}
\frac{\;d v^{\sigma}}{dt}\,+\,\Omega_{\mu\,\nu}^{\sigma}v^{\mu}v^{\nu} = 
{\mathscr T}_{\,\mu\,\nu}^{\,\sigma}v^{\mu}v^{\nu}, \label{3993}
\end{equation}
where ${v^{\sigma}}$ are the components of a tangent vector and ${t}$ is an affine parameter. The geodesics on ${S^3}$ in the
presence of torsion are thus different from the autoparallels with respect to the Weitzenb\"ock connection \cite{telepar}.
The straight lines with respect to ${\Omega_{\mu\,\nu}^{\sigma}}$  can be spirals or sinusoidal, and may appear ``bent'' to
our intuition. However, using the relation (\ref{W-con}) we can easily rewrite the above equation in terms of the
symmetric Levi-Civita connection ${\Gamma_{\mu\,\nu}^{\sigma}}$ as
\begin{equation}
\frac{\;d v^{\sigma}}{dt}\,+\,\Gamma_{\mu\,\nu}^{\sigma}v^{\mu}v^{\nu} =0\,. \label{3883}
\end{equation}
Evidently, the geodesics and autoparallels coincide for the Levi-Civita connection, which---in the absence of
torsion---corresponds to
${{\mathscr R}^{\,\alpha}_{\;\;\,\beta\,\gamma\,\delta}\not=0}$ and ${{\mathscr T}_{\,\alpha\,\beta}^{\,\gamma}=0}$.
The physical trajectory of the rotation governed by the above two equations in the presence of torsion
remain the same, however, because Eq.${\,}$(\ref{3883}) is simply a rewrite of Eq.${\,}$(\ref{3993}).

Given this equivalence, and the fact that torsion is an intrinsic property of the SU(2) manifold
[with the latter being isomorphic to the 3-sphere defined by Eq.${\,}$(\ref{nonoonpara})], it is
clear that Eq.${\,}$(\ref{dismes}) provides the most natural distance measure on SU(2), regardless of
its unusual appearance.

For the discussion in section V it is also important to recall here that the basis
${\left\{\,\beta_1({\bf p}),\;\beta_2({\bf p}),\;\beta_3({\bf p})\right\}}$ of ${\,T_{\bf p}S^3}$ at any ${p\in S^3}$
satisfies the anti-symmetric outer product
\begin{equation}
\frac{1}{2}\left[\,{\beta}_{\mu}\,,\;{\beta}_{\nu}\right]
\,=\,-\,\epsilon_{\mu\nu\rho}\,{\beta}_{\rho}\,, \label{commurela-alg-2}
\end{equation}
which, together with the symmetry of the inner product
\begin{equation}
\frac{1}{2}\left\{\,{\beta}_{\mu}\,,\;{\beta}_{\nu}\right\}
\,=\,-\,\delta_{\mu\nu}\,,
\end{equation}
leads to the algebra
\begin{equation}
{\beta}_{\mu}\,{\beta}_{\nu} \,=\,-\,\delta_{\mu\nu}
\,-\,\epsilon_{\mu\nu\rho}\,{\beta}_{\rho}\,.
\label{orig-alg}
\end{equation}
This is of course the familiar bivector subalgebra of the Clifford algebra ${{Cl}_{3,0}}$ of the orthogonal directions in the
physical space \cite{Clifford}\cite{Christian}. What may be less familiar is that this algebra also represents the local
structure---{\it i.e.,} the tangent space structure---of a unit parallelized 3-sphere. 

\section{Geodesic Distance on ${\rm SO(3)}$}

Our goal now is to obtain the geodesic distance on SO(3) by projecting the geodesic distance (\ref{dismes}) from
SU(2) onto SO(3) \cite{Abraham}. Physically this means considering
only those rotations that are insensitive to orientation-entanglement, or spinorial sign changes.
Now, as we saw in the previous section, SU(2) is homeomorphic to a set of unit quaternions and can be embedded into the
linear vector space ${{\rm I\!R}^4}$. A quaternion is thus a 4-vector within ${{\rm I\!R}^4}$, characterized by the
ordered and graded basis
\begin{equation}
\left\{1,\;{\bf e}_2\wedge{\bf e}_3,\;{\bf e}_3\wedge{\bf e}_1,\;{\bf e}_1\wedge{\bf e}_2\right\}.
\end{equation}
There are, however, twice as many elements in the set ${S^3}$ of all unit quaternions than there are points in the
configuration space SO(3) of all possible rotations in the physical space. This is because every pair of quaternions
constituting the antipodal points of ${S^3}$ represent one and the same rotation in ${{\rm I\!R}^3}$. This can be
readily confirmed by recalling how a quaternion and its antipode can rotate a bivector
${{\boldsymbol\beta}({\bf a})}$ about ${\bf a}$ to, say, a bivector ${{\boldsymbol\beta}({\bf a'})}$ about${\;\bf a'}$:
\begin{equation}
{\boldsymbol\beta}({\bf a'})\,=\,(+\,{\bf q})\,{\boldsymbol\beta}({\bf a})\,(+\,{\bf q})^{\dagger}
\,=\,(-\,{\bf q})\,{\boldsymbol\beta}({\bf a})\,(-\,{\bf q})^{\dagger}.
\end{equation}
Mathematically this equivalence is expressed by saying that ${S^3}$---or more precisely the group SU(2)---represents a
universal double covering of the rotation group SO(3) \cite{Analysis}.
The configuration space of all possible rotations in the physical space is therefore obtained by identifying the
antipodal points of ${S^3}$---{\it i.e.}, by identifying every point ${+\,{\bf q}\in S^3}$ with its antipode ${-\,{\bf q}\in S^3}$.
The space that results from this identification is a real projective space, ${{\rm I\!R}{\rm P}^3}$, which is simply
the set of all lines through the origin of ${{\rm I\!R}^4}$. There are thus precisely two preimages in ${S^3}$, namely
${+\,{\bf q}}$ and ${-\,{\bf q}}$, corresponding to each rotation ${R_{\bf q}}$ in SO(3). As a result, ${S^3}$
is realized as a fiber bundle over ${{\rm I\!R}{\rm P}^3}$ with each fiber consisting of exactly two points:
\begin{equation}
S^3/\{-1,\,+1\}\,\approx\,{\rm I\!R}{\rm P}^3.
\end{equation}
In other words, ${{\rm I\!R}{\rm P}^3}$ is the quotient of ${S^3}$ by the map ${{\bf q}\mapsto -\,{\bf q}}$, which we
shall denote by ${\varphi:S^3\rightarrow{\rm I\!R}{\rm P}^3}$.

Although this quotient map renders the topologies of the spaces ${S^3}$ and ${{\rm I\!R}{\rm P}^3}$ quite distinct from one
another (for example the space ${S^3}$ is simply-connected, whereas the space ${{\rm I\!R}{\rm P}^3}$ is connected but not
simply-connected), it leaves their Lie algebra structure ({\it i.e.}, their tangent space structure) unaffected. Therefore
it is usually not possible to distinguish the signatures of the spinorial group SU(2) from those of the tensorial group SO(3)
by local measurements alone. In fact the space ${{\rm I\!R}{\rm P}^3}$---being homeomorphic to a Lie group---is
just as parallelizable as the space ${S^3}$. In more familiar terms this means that it is impossible to tell by local
observations alone whether a given object has undergone even number of ${2\pi}$ rotations prior to observation or odd number
of ${2\pi}$ rotations \cite{Hartung}.

We are, however, interested precisely in distinguishing the global properties of ${S^3}$ from those of
${{\rm I\!R}{\rm P}^3}$ by means of local, albeit relative measurements \cite{Weingard}. Fortunately, it turns out to be possible
to distinguish and compare the geodesic distances on the spaces ${S^3}$ and ${{\rm I\!R}{\rm P}^3}$ induced by the
Euclidean metric ${\delta_{\mu\nu}}$ of the parallelized ${S^3}$. This is possible because, for any Lie group, such as
SO(3), with a left-invariant metric coinciding at the identity with a left-invariant metric on
its universal covering group, such as SU(2), the geodesics on the covering group are simply the horizontal lifts of the geodesics
on that group \cite{Abraham}. In other words, we can induce a metric on the base space ${{\rm I\!R}{\rm P}^3}$ by projecting the
Euclidean metric of the total space ${S^3}$ by means of the quotient map ${\varphi:S^3\rightarrow{\rm I\!R}{\rm P}^3}$.
This induced metric then provides a measure of the respective geodesic distances
within the manifolds ${S^3}$ and ${{\rm I\!R}{\rm P}^3}$.

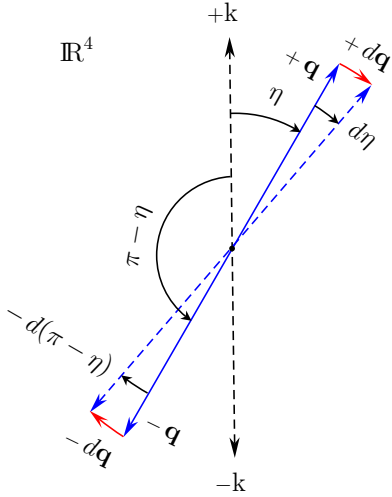
\begin{figure}
\hrule
\scalebox{0.7}{
\begin{pspicture}(-6.0,-5.2)(5.0,5.0)

\begin{rotate}{0}
\psline[linewidth=0.3mm,linestyle=dashed,arrowinset=0.3,arrowsize=3pt 3,arrowlength=2]{<->}(0.0,-4.25)(-0.1,3.75)
\end{rotate}

\begin{rotate}{-30}
\psline[linewidth=0.3mm,linecolor=blue,arrowinset=0.3,arrowsize=3pt 3,arrowlength=2]{<->}(0.0,-4.45)(0.0,3.75)
\end{rotate}

\begin{rotate}{-40}
\psline[linewidth=0.3mm,linestyle=dashed,linecolor=blue,arrowinset=0.3,arrowsize=3pt 3,arrowlength=2]{<->}(-0.1,-4.5)(0.0,3.7)
\end{rotate}

\psarc[linewidth=0.3mm,arrowinset=0.3,arrowsize=3pt 3,arrowlength=1]{<-}(-0.2,0.0){2.3}{57.5}{92.5}

\psarc[linewidth=0.3mm,arrowinset=0.3,arrowsize=3pt 3,arrowlength=1]{->}(-0.2,-0.4){1.5}{93.7}{236}

\psarc[linewidth=0.3mm,linecolor=red,arrowsize=3pt 3,arrowlength=2]{<-}(0.1,0.0){3.65}{51}{62.7}

\psarc[linewidth=0.3mm,linecolor=red,arrowsize=3pt 3,arrowlength=2]{<-}(0.1,0.0){4.55}{227.5}{237.7}

\psarc[linewidth=0.3mm,arrowinset=0.3,arrowsize=3pt 3,arrowlength=1]{<-}(-0.2,0.0){2.87}{47}{58.5}

\psarc[linewidth=0.3mm,arrowinset=0.3,arrowsize=3pt 3,arrowlength=1]{<-}(-0.2,0.0){3.45}{231}{241}

\uput[90](-0.45,3.85){\large +{\Large k}}

\uput[90](-0.3,-5.1){\large ${-}${\Large k}}

\begin{rotate}{-28}
\uput[90](-0.6,3.0){\large +{\Large ${\,{\bf q}}$}}
\end{rotate}

\begin{rotate}{-26}
\uput[90](0.1,-4.45){\large ${-}${\Large ${\,{\bf q}}$}}
\end{rotate}

\begin{rotate}{-28}
\uput[90](0.2,3.65){\large +{\Large ${\,d{\bf q}}$}}
\end{rotate}

\begin{rotate}{-26}
\uput[90](-1.22,-5.5){\large ${-}${\Large ${\,d{\bf q}}$}}
\end{rotate}

\begin{rotate}{-280}
\rput[-30](-0.47,2.45){\Large ${\pi-\eta}$}
\end{rotate}

\begin{rotate}{-15}
\uput[90](-0.67,2.2){\Large ${\eta}$}
\end{rotate}

\begin{rotate}{-30}
\uput[90](0.4,2.0){\Large ${d\eta}$}
\end{rotate}

\begin{rotate}{-30}
\uput[90](-2.75,-4.33){\Large ${-\,d(\pi-\eta)}$}
\end{rotate}

\pscircle*(-1.03,-0.27){0.055}

\put(-4.3,3.3){{\Large ${{\rm I\!R}^4}$}}

\end{pspicture}}
\vspace{0.25cm}
\hrule
\caption{Projecting the metric tensor from ${T_{\bf q}S^3}$ to ${T_{[{\bf q}]}{\rm I\!R}{\rm P}^3}$.\break}
\label{fig-notsoso}
\vspace{0.3cm}
\hrule
\end{figure}

To deduce this metric, let us recall the definitions of the tangent space ${T_{\bf q}S^3}$ and the
tangent bundle ${TS^3}$ from the previous section [cf. Eqs.${\,}$(\ref{tanspace}) and (\ref{tanbundle})]. Under the
map ${\varphi:S^3\rightarrow{\rm I\!R}{\rm P}^3}$ the tangent bundle ${T{\rm I\!R}{\rm P}^3}$ is then the
quotient of the tangent bundle ${TS^3}$ by the involution
\begin{equation}
({\bf q},\,{\bf t_q})\,\mapsto\,(-\,{\bf q},\,-\,{\bf t_q}).
\end{equation}
In other words, ${({\bf q},\,{\bf t_q})\in T_{\bf q}S^3}$ and ${(-\,{\bf q},\,-\,{\bf t_q})\in T_{-\,{\bf q}}S^3}$ have the same
image in the tangent space
${T_{[\,{\bf q}\,]}{\rm I\!R}{\rm P}^3}$ under the derivative map ${d\varphi:TS^3\rightarrow T{\rm I\!R}{\rm P}^3}$, where
${[\,{\bf q}\,]\in{\rm I\!R}{\rm P}^3}$ is a line through ${\pm\,{\bf q}}$ in ${{\rm I\!R}^4}$. Since the quotient map
${\varphi({\bf q})}$ is a local diffeomorphism, its derivative map
\begin{equation}
d\varphi({\bf q}):\,T_{\bf q}S^3\rightarrow T_{[\,{\bf q}\,]}{\rm I\!R}{\rm P}^3 \label{dermappa}
\end{equation}
is an isomorphism for every ${{\bf q}\in S^3}$. Thus the tangent space ${T_{[\,{\bf q}\,]}{\rm I\!R}{\rm P}^3}$ can be
identified with the space of pairs:
\begin{equation}
\!\left\{({\bf q},\,{\bf t_q}),\,(-\,{\bf q},\,-\,{\bf t_q})\Big|\,
{\bf q},\,{\bf t_q}\in {\rm I\!R}^4\!,\,||{\bf q}||=1,\,{\bf q}\cdot{\bf t}_{\bf q}^{\dagger}=0 \right\}\!. \label{tanpac2}
\end{equation}
One such pair is depicted in Fig.${\,}$\ref{fig-notsoso}, with the spinor ${+{\bf q}}$ making an angle ${\eta}$ with a
reference spinor ${+{\bf k}}$. It is easy to see from this figure that at one point of ${S^3}$ we have the differential relation
\begin{align}
d{\bf q}\,=\,{\bf t}_{\bf q}\;d\eta\,,
\end{align}
whereas at its antipodal point we have the relation 
\begin{equation}
d{\bf q}\,=\,-\,{\bf t}_{\bf q}\;d(\pi-\eta)\,.
\end{equation}
Identifying ${{\bf t}_{\bf q}}$ with ${-\,{\bf t}_{\bf q}}$ thus amounts to identifying
${d\eta}$ with ${d(\pi-\eta)}$:
\begin{equation}
d\eta\,=\,d(\pi-\eta)\,.
\end{equation}
This identification can be effected by the following change in the variable ${\eta}$:
\begin{equation}
\eta\,\mapsto\,\alpha\,=\,-\,\pi\,\cos\eta\,.
\end{equation}
Consequently, the infinitesimal measure of distance on ${S^3}$ that projects down to ${{\rm I\!R}{\rm P}^3}$ as a result of
the quotient map ${\varphi:S^3\rightarrow{\rm I\!R}{\rm P}^3}$ is
\begin{equation}
\pi\,\sin\eta\,d\eta\,\mapsto\,d\alpha\,,
\end{equation}
where ${d\alpha}$ is the infinitesimal measure of distance on ${{\rm I\!R}{\rm P}^3}$.
Since ${\sin(\pi-\eta)=\sin\eta}$, it is easy to see that the pair
\begin{equation}
\{\,\pi\,\sin\eta\,d\eta,\;\pi\,\sin(\pi-\eta)\,d\eta\,\}
\end{equation}
has the same image ${d\alpha}$ in ${T_{[\,{\bf q}\,]}{\rm I\!R}{\rm P}^3}$. The corresponding finite interval
${[0,\,\eta_{{\bf a}{\bf b}}]}$ then projects down to ${{\rm I\!R}{\rm P}^3}$ as
\begin{equation}
\pi\int_{0}^{\eta_{{\bf a}{\bf b}}}\sin\eta\,d\eta\,\;\longmapsto
\int_{0}^{2\eta_{{\bf a}{\bf b}}}d\alpha\,,
\end{equation}
where the factor of 2 appearing in the upper limit on the RHS reflects the universal double covering of
${{\rm I\!R}{\rm P}^3}$ by ${S^3}$. These integrals can now be evaluated separately for the intervals
${0 \leq \eta_{{\bf a}{\bf b}} \leq \pi}$ and ${\pi \leq \eta_{{\bf a}{\bf b}} \leq 2\pi}$ to give
\begin{align}
-\,\cos\eta_{{\bf a}{\bf b}}\,\mapsto\,&
\begin{cases}
-\,1\,+\,\frac{2}{\pi}\,\eta_{{\bf a}{\bf b}}
\,\;\;\;{\rm if} & 0 \leq \eta_{{\bf a}{\bf b}} \leq \pi \\
\\
+\,3\,-\,\frac{2}{\pi}\,\eta_{{\bf a}{\bf b}}
\,\;\;\;{\rm if} & \pi \leq \eta_{{\bf a}{\bf b}} \leq 2\pi \label{whrthichr}
\end{cases} \\
&=\,-\,\cos\alpha_{{\bf a}{\bf b}}\,, \notag
\end{align}
which---as we shall soon confirm---is an expression of the fact that the map ${\varphi:S^3\rightarrow{\rm I\!R}{\rm P}^3}$ is a
{\it local} isometry [cf. Eqs.${\,}$(\ref{troeeq}) to (\ref{91913})].
Thus the geodesic distance projected from the distance (\ref{dismes}) in ${S^3}$
onto ${{\rm I\!R}P^3}$ is given by 
\begin{align}
{\mathscr D}({\bf a},\,{\bf b})&=
\begin{cases}
-\,1\,+\,\frac{2}{\pi}\,\eta_{{\bf a}{\bf b}}
\;\;\;{\rm if} & \! 0 \leq \eta_{{\bf a}{\bf b}} \leq \pi \\
\\
+\,3\,-\,\frac{2}{\pi}\,\eta_{{\bf a}{\bf b}}
\;\;\;{\rm if} & \! \pi \leq \eta_{{\bf a}{\bf b}} \leq 2\pi\,, 
\end{cases}  \label{dismee}
\end{align}
where ${\eta_{{\bf a}{\bf b}}}$ is half of the rotation angle ${\psi}$, just as in (\ref{dismes}).

Needless to say, this measure of geodesic distance can also be obtained directly from the geometry of SO(3) itself
\cite{Spinor}\cite{Huynh}.
Since it is a compact Lie group, SO(3) has a natural Riemannian metric---{\it i.e.}, an inner product on its tangent space
${T_p\,{\rm SO(3)}}$ at every point ${p}$ \cite{Analysis}. At the identity of SO(3) this tangent space is isomorphic to the Lie
algebra ${so(3)}$ of skew-symmetric matrices of the form
\begin{equation}
\left(\begin{matrix} \; 0 \; & \; -a_3 \; & \; +a_2 \; \\ \\
                     \; +a_3 \; & \; 0 \; & \; -a_1 \; \\ \\
                     \; -a_2 \; & \; +a_1 \; & \; 0 \;
\end{matrix}\right)\!, \label{metss}
\end{equation}
where ${{\bf a}\in{\rm I\!R}^3}$. These elements of Lie algebra so(3) can be represented also by unit bivectors of the form
${{\boldsymbol\xi}({\bf a})\in S^2}$ satisfying the anti-symmetric outer product as before,
\begin{equation}
\frac{1}{2}\left[\,{\xi}_{\mu}\,,\;{\xi}_{\nu}\right]
\,=\,-\,\epsilon_{\mu\nu\rho}\,{\xi}_{\rho}\,, \label{commurela-alg-1}
\end{equation}
with the basis bivectors ${\{\;{\xi}_1({\bf p}),
\;{\xi}_2({\bf p}),\;{\xi}_3({\bf p})\,\}}$ spanning the tangent spaces at every point
${p\in{\rm SO(3)}}$. These basis bivectors, however, respect a general symmetric product
\begin{equation}
\frac{1}{2}\left\{\,{\xi}_{\mu}\,,\;{\xi}_{\nu}\right\}
\,=\,-\,J_{\mu\nu}\,, \label{anti-commurela-alg}
\end{equation}
with the metric tensor ${J}$ now providing a non-Euclidean inner product on SO(3):
\begin{align}
\langle{\xi}_{\mu},\,{\xi}^{\dagger}_{\nu}\rangle
\,=\,J_{\mu}^{\;\,\rho}{\beta}_{\rho}\cdot{\beta}^{\dagger}_{\nu}\,=\,J_{\mu}^{\;\,\rho}\delta_{\rho\nu}
\,=\,J_{\mu\nu}\,.\label{orige-onoflat}
\end{align}
This metric can be verified by multiplying two matrices of the form (\ref{metss}), with the usual Euclidean inner
product emerging as one-half of the trace of the product matrix.
Physically ${J}$ can be interpreted as a moment of inertia tensor of a freely rotating asymmetrical top \cite{Abraham}.
For a spherically symmetrical top ${J_{\mu\nu}}$ reduces to ${\delta_{\mu\nu}}$ and the Euclidean
inner product (\ref{flat-me}) is recovered \cite{Kambe}. In general the orientation of the top
traces out a geodesic distance in SO(3) with respect to its inertia tensor ${J}$ \cite{Abraham}\cite{Arnold}.

\begin{figure}
\hrule
\scalebox{0.77}{
\begin{pspicture}(4.5,-1.0)(5.0,5.9)
\psset{xunit=0.5mm,yunit=4cm}
\psaxes[axesstyle=frame,tickstyle=full,ticksize=0pt,dx=90\psxunit,Dx=180,dy=1\psyunit,Dy=+2,Oy=-1](0,0)(180,1.0)
\psline[linewidth=0.3mm,arrowinset=0.3,arrowsize=2pt 3,arrowlength=2]{->}(0,0.5)(190,0.5)
\psline[linewidth=0.2mm]{-}(45,0)(45,1)
\psline[linewidth=0.2mm]{-}(90,0)(90,1)
\psline[linewidth=0.2mm]{-}(135,0)(135,1)
\psline[linewidth=0.35mm,arrowinset=0.3,arrowsize=2pt 3,arrowlength=2]{->}(0,0)(0,1.2)
\psline[linewidth=0.35mm,linestyle=dashed,linecolor=blue]{-}(0,0)(90,1)
\psline[linewidth=0.35mm,linestyle=dashed,linecolor=blue]{-}(90,1)(180,0)
\put(2.1,-0.38){${90}$}
\put(6.5,-0.38){${270}$}
\put(-0.63,3.92){${+}$}
\put(-0.6,5.0){{\large ${\mathscr D}$}${({\bf a},\,{\bf b})}$}
\put(-0.38,1.93){${0}$}
\put(9.65,1.85){\large ${\eta_{{\bf a}{\bf b}}}$}
\psplot[linewidth=0.35mm,linecolor=red]{0.0}{180}{x dup cos exch cos mul 1.0 mul neg 1 add}
\end{pspicture}}
\hrule
\caption{Comparison of the geodesic distances on ${\rm SU(2)}$ and ${\rm SO(3)}$ as functions of
half of the rotation angle ${\psi}$. The dashed lines depict the geodesic distances on ${\rm SO(3)}$,
whereas the red\break curve depicts their horizontal lift to the covering group ${\rm SU(2)}$.}
\vspace{0.3cm}
\label{fig-55}
\hrule
\end{figure}
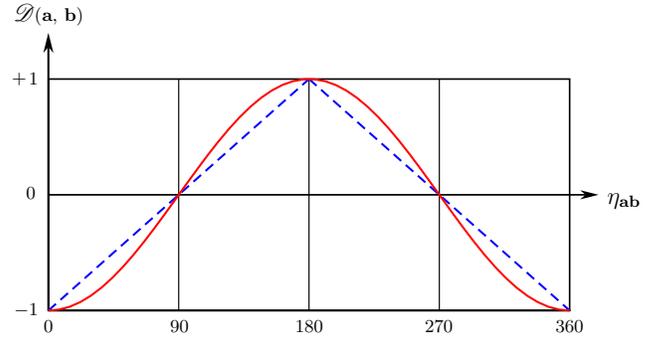

To calculate the length of this distance, recall that the inner product in the Riemannian structure
provides an infinitesimal length on the tangent space so that the length of a curve can be obtained by
integration along the curve \cite{Analysis}. Then the shortest path---{\it i.e.}, the geodesic
distance---from the identity of SO(3) to another point can be obtained by means of the exponential map
\begin{equation}
\exp\,:\,T_e\,{\rm SO(3)}\approx so(3)\rightarrow{\rm SO(3)}, 
\end{equation}
which maps the line ${{\boldsymbol\xi}({\bf a})\,t_{\bf a}}$
in the tangent space ${T_e\,{\rm SO(3)}}$ at the origin ${e}$ of SO(3) onto the group SO(3) such that
\begin{equation}
{\boldsymbol\xi}({\bf a})\mapsto
\exp{\left\{{\boldsymbol\xi}({\bf a})\right\}}\!
:=\exp{\left\{{\boldsymbol\xi}({\bf a})\,t_{\bf a}\right\}}\bigg|_{t_{\bf a}=\,1}\,, \label{denofi}
\end{equation}
where ${{\boldsymbol\xi}({\bf a})\in so(3)}$ with
${{\boldsymbol\xi}({\bf a}){\boldsymbol\xi}^{\dagger}\!({\bf a})=1}$, and
\begin{equation}
t_{\bf a}\,=\,
\begin{cases}
-\,1\,+\,\frac{1}{\pi}\,\psi_{\bf a}\,\;\;\;\;{\rm if} & 0 \leq \psi_{\bf a} \leq 2\pi \\
\\
+\,3\,-\,\frac{1}{\pi}\,\psi_{\bf a}\,\;\;\;\;{\rm if} & 2\pi \leq \psi_{\bf a} \leq 4\pi\,. \label{tdein-2}
\end{cases}
\end{equation}
Thus ${t_{\bf a}}$ takes values from the interval ${[-1,\,+1]}$, with ${\psi_{\bf a}}$ being the angle of rotation as before.
More importantly, ${t_{\bf a}}$ provides a measure of a geodesic distance between the identity of SO(3) and the element
\begin{equation}
{\cal R}_{\bf a}:=\exp{\left\{{\boldsymbol\xi}({\bf a})\,t_{\bf a}\right\}}\in {\rm SO(3)}\,.
\end{equation} 
Therefore, we can use this exponential map to define a bi-invariant distance measure on SO(3) as
\begin{equation}
{\mathscr D}: {\rm SO(3)}\times{\rm SO(3)} \rightarrow {\rm I\!R}
\end{equation}
such that
\begin{equation}
{\mathscr D}({\bf a},\,{\bf b})
=\left|\left|\,\log\left\{{\cal R}^{\,}_{\bf a}\,{\cal R}^{\dagger}_{\bf b}\right\}\right|\right|\!.
\end{equation}
This distance measure calculates the amount of rotation required to bring ${{\cal R}_{\bf a}\in{\rm SO(3)}}$ to align
with ${{\cal R}_{\bf b}\in{\rm SO(3)}}$ by finding ${{\cal R}_{\bf ab}\in{\rm SO(3)}}$ such
that ${{\cal R}_{\bf a}={\cal R}_{\bf ab}\,{\cal R}_{\bf b}}$, because then we have
${{\cal R}_{\bf ab}={\cal R}^{\,}_{\bf a}\,{\cal R}^{\dagger}_{\bf b}}$
since ${{\cal R}^{\,}_{\bf b}\,{\cal R}^{\dagger}_{\bf b}=1}$.
Thus, in view of definition (\ref{denofi}), we arrive at the following distance between
${{\cal R}_{\bf a}}$ and ${{\cal R}_{\bf b}}$ (see also Fig.${\,}$11 of Ref.${\,}$\cite{Hestenes}):
\begin{align}
{\mathscr D}({\bf a},\,{\bf b})=
\begin{cases}
-\,1\,+\,\frac{1}{\pi}\,\psi_{\bf ab}\,\;\;\;\;{\rm if} & 0 \leq \psi_{\bf ab} \leq 2\pi \\
\\
+\,3\,-\,\frac{1}{\pi}\,\psi_{\bf ab}\,\;\;\;\;{\rm if} & 2\pi \leq \psi_{\bf ab} \leq 4\pi\,. \label{tdein-1}
\end{cases}
\end{align}
But since ${\psi_{\bf ab}=2\,\eta_{\bf ab}}$, this distance is equivalent to the one obtained in (\ref{dismee}).
In Fig.${\,}$\ref{fig-55}
both distance functions (\ref{dismes}) and (\ref{dismee}) are plotted as functions of the angle ${\eta_{\bf ab}}$.

Next, using the inner product (\ref{orige-onoflat}) and the fact that orientations of a rotating body trace out a geodesic
in SO(3) with respect to its inertia tensor, we can rewrite the above measure in terms of the metric tensor ${J}$ as 
\begin{align}
-\,J_{\mu\nu}\,a_{\mu}\,b_{\nu}&=:\,-\,\cos\alpha_{{\bf a}{\bf b}}\, \notag \\
&=
\begin{cases}
-\,1\,+\,\frac{2}{\pi}\,\eta_{{\bf a}{\bf b}}
\;\;\;\;{\rm if} & 0 \leq \eta_{{\bf a}{\bf b}} \leq \pi \\
\\
+\,3\,-\,\frac{2}{\pi}\,\eta_{{\bf a}{\bf b}}
\;\;\;\;{\rm if} & \pi \leq \eta_{{\bf a}{\bf b}} \leq 2\pi, \label{equawhichr}
\end{cases}
\end{align}
with SO(3) counterpart of the inner product (\ref{conpat}) being
\begin{align}
\langle\,\{a_{\mu}\;{\xi}_{\mu}({\bf p})\},\;\{b_{\nu}\;{\xi}_{\nu}({\bf q})\}\,\rangle&=
\langle\,\{a_{\mu}\;{\xi}_{\mu}({\bf q})\},\;\{b_{\nu}\;{\xi}_{\nu}({\bf q})\}\,\rangle\, \notag \\
&=-\,\langle\;\xi_{\mu}({\bf q}),\;\xi^{\dagger}_{\nu}({\bf q})\,\rangle\;a_{\mu}\,b_{\nu}\, \notag \\
&=-\,J_{\mu\nu}\,a_{\mu}\,b_{\nu}\, \notag \\
&=\,-\,\cos\alpha_{{\bf a}{\bf b}}\,.
\end{align}
Here ${\eta_{{\bf a}{\bf b}}}$ is the angle between tangent bivectors ${a_{\mu}{\beta}_{\mu}({\bf q})}$ and
${b_{\nu}{\beta}_{\nu}({\bf q})}$ within ${T_{\bf q}S^3}$, whereas ${\alpha_{{\bf a}{\bf b}}}$ is the angle
between tangent bivectors ${a_{\mu}{\xi}_{\mu}({\bf q})}$ and ${b_{\nu}{\xi}_{\nu}({\bf q})}$ within
${T_{[{\bf q}]}{\rm I\!R}{\rm P}^3}$. Consequently, using the algebraic identity
\begin{equation}
{\xi}_{\mu}\,{\xi}_{\nu} \,=\,
\frac{1}{2}\left\{\,{\xi}_{\mu}\,,\;{\xi}_{\nu}\right\}\,+\,
\frac{1}{2}\left[\,{\xi}_{\mu}\,,\;{\xi}_{\nu}\right] \label{identcommure}
\end{equation}
and the equations (\ref{commurela-alg-1}) and (\ref{anti-commurela-alg}), we arrive at the algebra
\begin{equation}
{\xi}_{\mu}\,{\xi}_{\nu}\,=\,-\,J_{\mu\nu}\,-\,\epsilon_{\mu\nu\rho}\,{\xi}_{\rho}\,. \label{norig-alg}
\end{equation}
This induced algebra is then the SO(3) counterpart of the algebra (\ref{orig-alg}). Note that
it coincides with the algebra (\ref{orig-alg}) at the identity ({\it i.e.}, when
${\psi_{{\bf a}{\bf b}}=\pi}$ and ${\psi_{{\bf a}{\bf b}}=3\pi}$).

It is important to recognize here that the quotient map ${\varphi:S^3\rightarrow{\rm I\!R}{\rm P}^3}$ we considered above is a
{\it local} isometry. The infinitesimal map ${d\varphi({\bf q}):\,T_{\bf q}S^3\rightarrow T_{[\,{\bf q}\,]}{\rm I\!R}{\rm P}^3}$
is thus an isometry. The inner product in ${T_{\bf q}S^3}$ defined by (\ref{orige-flat}) is therefore locally preserved under the
action of ${\varphi}$:
\begin{align}
\langle\,\{a_{\mu}\,{\beta}_{\mu}({\bf q})&\},\;\{b_{\nu}\,{\beta}_{\nu}({\bf q})\}\,\rangle \notag \\
&=\,\langle\,\{a_{\mu}\,{\xi}_{\mu}[\varphi({\bf q})]\},\;\{b_{\nu}\,{\xi}_{\nu}[\varphi({\bf q})]\}\,\rangle\,. \label{troeeq}
\end{align}
Note, however, that this equality between the metrics in ${T_{\bf q}S^3}$ and ${T_{[\,{\bf q}\,]}{\rm I\!R}{\rm P}^3}$
holds only for the values ${\varphi({\bf q})={\bf q}}$, ${\varphi({\bf q})=-{\bf q}}$, ${\varphi({\bf q})={\bf t_q}}$, and
${\varphi({\bf q})=-{\bf t_q}}$, which can be verified from the definition (\ref{tanpac2}) of the tangent space
${T_{[\,{\bf q}\,]}{\rm I\!R}{\rm P}^3}$ with the help of Fig.${\,}$\ref{fig-notsoso}. For definiteness, let the quaternions
${\bf q}$ shown in Fig.${\,}$\ref{fig-notsoso} be parameterized as
\begin{align}
{\bf q_a}\,&=\,\cos\eta_{\bf a}\,+\,{\boldsymbol\beta}({\bf c})\,\sin\eta_{\bf a}\,, \notag \\
{\bf q_b}\,&=\,\cos\eta_{\bf b}\,+\,{\boldsymbol\beta}({\bf c})\,\sin\eta_{\bf b}\,,
\end{align}
{\it etc.}, along with its tangent quaternions parameterized as
\begin{align}
{\bf t_{q_a}}\,&=\,-\sin\eta_{\bf a}\,+\,{\boldsymbol\beta}({\bf c})\,\cos\eta_{\bf a}\,, \notag \\
{\bf t_{q_b}}\,&=\,-\sin\eta_{\bf b}\,+\,{\boldsymbol\beta}({\bf c})\,\cos\eta_{\bf b}\,,
\end{align}
{\it etc.}, where ${{\bf c}={\bf a}\times{\bf b}/|{\bf a}\times{\bf b}|}$.
Then the angle between ${\bf q_a}$ and ${\bf q_b}$ (as well as between ${\bf t_{q_a}}$ and ${\bf t_{q_b}}$) is given by
\begin{equation}
\cos\eta_{\bf ab}\,=\,|\,{\bf q_a}\cdot{\bf q^{\dagger}_b}|\,=\,|\cos\eta_{\bf a}\cos\eta_{\bf b}
\,+\,\sin\eta_{\bf a}\sin\eta_{\bf b}|\,. \label{91913}
\end{equation}
From Fig.${\,}$\ref{fig-notsoso} it is now easy to see that the projection
${\varphi({\bf q})={\bf q}}$ corresponds to ${\eta_{{\bf a}{\bf b}}=0}$ and ${\eta_{{\bf a}{\bf b}}=2\pi}$, whereas
the projection ${\varphi({\bf q})=-{\bf q}}$ corresponds to ${\eta_{{\bf a}{\bf b}}=\pi}$. Two further possibilities,
${\varphi({\bf q})={\bf t_q}}$ and ${\varphi({\bf q})=-{\bf t_q}}$, which correspond to the projections
${({\bf q},\,{\bf t_q})\mapsto({\bf t_q},\,-{\bf q})}$ and ${({\bf q},\,{\bf t_q})\mapsto(-{\bf t_q},\,{\bf q})}$,
are also permitted by the definition (\ref{tanpac2}) of ${T_{[\,{\bf q}\,]}{\rm I\!R}{\rm P}^3}$, because
${{\rm T}\,{\rm I\!R}{\rm P}^3}$ is invariant under the rotation of ${\bf q}$ by ${\pi/2}$. They amount to swaps
in the roles played by ${\bf q}$ and ${\bf t_q}$.
From Fig.${\,}$\ref{fig-notsoso} it is easy to see that these possibilities correspond to
${\eta_{{\bf a}{\bf b}}=\pi/2}$ and ${\eta_{{\bf a}{\bf b}}=3\pi/2}$.

The argument above shows that the equality (\ref{troeeq}) holds only for the angles ${\eta_{{\bf a}{\bf b}}=0}$, ${\pi/2}$,
${\pi}$, ${3\pi/2}$, and ${2\pi}$. That is to say, the local isometry (\ref{troeeq}) under the map
${\varphi:S^3\rightarrow{\rm I\!R}{\rm P}^3}$ requires that the geodesic distances on ${S^3}$ and ${{\rm I\!R}{\rm P}^3}$ agree
{\it only} in the infinitesimal neighborhoods of the angles ${\eta_{{\bf a}{\bf b}}=0}$, ${\pi/2}$, ${\pi}$, ${3\pi/2}$,
and ${2\pi}$. They are not preserved globally. This is confirmed in Fig.${\,}$\ref{fig-55}, which is a plot of the distance
functions (\ref{dismes}) and (\ref{dismee}). The geodesic distances on SU(2) and SO(3) are evidently different for the angles
other than ${\eta_{{\bf a}{\bf b}}=0}$, ${\pi/2}$, ${\pi}$, ${3\pi/2}$, and ${2\pi}$. The underlying reason for the difference
in the shape of the two geodesic distances is the difference in the topologies of ${S^3}$ and ${{\rm I\!R}{\rm P}^3}$.
While ${S^3}$ is simply-connected, ${{\rm I\!R}{\rm P}^3}$ is connected but not simply-connected. As a result the map
${\varphi:S^3\rightarrow{\rm I\!R}{\rm P}^3}$ is only a local and not a global isometry. Moreover, while both ${S^3}$ and
${{\rm I\!R}{\rm P}^3}$ are parallelized by the basis ${\{\beta_{\mu}\}}$ and ${\{\xi_{\nu}\}}$, respectively, the torsion
within these manifolds is different. And it is this difference in torsion that is reflected in the shapes of the geodesics
depicted in Fig.${\,}$\ref{fig-55}. In the appendices A and C below we spell out the parallelization process in more detail
and bring out the crucial role played by torsion in these manifolds.

To sum up, we started out with a parallelized 3-sphere representing SU(2) and then identified each of
its points with its antipodal point to obtain a parallelized ${{\rm I\!R}{\rm P}^3}$ representing SO(3).
But in doing so we lost the following spinorial symmetry satisfied by ${{\bf q}(\psi,\,{\bf r})}$, as
described by equations (\ref{clacct}) and (\ref{non-clacct}) in the Introduction:
\begin{equation}
{\bf q}(\psi+2\kappa\pi,\,{\bf r})\,=\,-\,{\bf q}(\psi,\,{\bf r})\,\;\;\text{for}\;\,\kappa=1,3,5,7,\dots
\end{equation}
In other words, by identifying the antipodal points of ${S^3}$ we lost the sensitivity of ${{\bf q}(\psi,\,{\bf r})}$
to spinorial sign changes.
As a result, ${{\bf q}(\psi,\,{\bf r})}$ constituting ${{\rm I\!R}{\rm P}^3}$ represents the state of a rotating
body that returns to itself after {\it any} multiple of ${2\pi}$ rotation:
\begin{equation}
{\bf q}(\psi+2\kappa\pi,\,{\bf r})\,=\,+\,{\bf q}(\psi,\,{\bf r})\,\;\;\text{for}\;\,\text{any}\;\,\kappa=0,1,2,3,\dots
\end{equation}
A measurable difference between these two possibility is then captured by the geodesic distances on the manifolds
SU(2) and SO(3), as depicted, for example, in Fig.${\,}$\ref{fig-55}.

\section{Proposed Experiment}

We now sketch a classical experiment which could, in principle, distinguish the geodesic distance on SU(2) from that on SO(3),
as calculated in the previous sections (cf. Fig.${\,}$\ref{fig-55}). It can be performed either in the outer space or in a
terrestrial laboratory. In the latter case the effects of gravity and air resistance would complicate matters, but it may
be possible to choose experimental parameters sufficiently carefully to compensate for such effects.

With this assumption, consider a ``bomb'' made out of a hollow toy ball of diameter, say, three centimeters.
The thin hemispherical shells of uniform density that make up the ball are snapped together at their rims
in such a manner that a slight increase in temperature would pop the ball open into its two constituents
with considerable force. A small lump of density much greater than the density of the ball
is attached on the inner surface of each shell at a random location, so that, when the ball pops open,
not only would the two shells propagate with equal and opposite linear momenta orthogonal to their common
plane, but would also rotate with equal and opposite spin momenta about a random axis in space. The volume of
the attached lumps can be as small as a cubic millimeter, whereas their mass can be comparable to the mass of
the ball. This will facilitate some ${10^6}$ possible spin directions for the two shells, whose outer surfaces
can be decorated with colors to make their rotations easily detectable.

Now consider a large ensemble of such balls, identical in every respect except for the relative locations of the
two lumps (affixed randomly on the inner surface of each shell). The balls are then placed over a heater---one at a
time---at the center of the experimental setup \cite{Peres}, with the common plane of their shells held perpendicular
to the horizontal direction of the setup. Although initially at rest, a slight increase in temperature of each
ball will eventually eject its two shells towards the observation stations, situated at a chosen distance in the
mutually opposite directions. Instead of selecting the directions ${\bf a}$ and ${\bf b}$ for observing spin
components, however, one or more contact-less rotational motion sensors---capable of determining the precise
direction of rotation---are placed near each of the two stations, interfaced with a computer. These
sensors will determine the exact direction of the spin angular momentum ${{\bf s}^k}$
(or ${-{\bf s}^k}$) for each shell in a given explosion, without disturbing them otherwise so that their
total angular momentum would remain zero, at a designated distance from
the center. The interfaced computers can then record this data, in the form of a 3D map of all such directions,
at each observation station.

Once the actual directions of the angular momenta for a large ensemble of shells on both sides are fully recorded,
the two computers are instructed to randomly choose a pair of
reference directions, say ${\bf a}$ for one station and ${\bf b}$ for
the other station---from the two 3D maps of already existing data---and then calculate the corresponding pair of numbers
${{sign}\,({\bf s}^k\cdot{\bf a})=\pm\,1}$ and ${{sign}\,(\,-\,{\bf s}^k\cdot{\bf b})=\pm\,1}$.
The correlation function for the bomb fragments can then be calculated as
\begin{equation}
{\cal E}({\bf a},\,{\bf b})=\!\!\lim_{\,n\,\gg\,1}\!\left[\frac{1}{n}\!\sum_{k\,=\,1}^{n}
\{{sign}\,(+{\bf s}^k\cdot{\bf a})\}\,
\{{sign}\,(-{\bf s}^k\cdot{\bf b})\}\right]\!,\label{correlations}
\end{equation}
where ${n}$ is the total number of experiments performed.

In the next section we shall see how this correlation function provides a measure of geodesic distances on the manifolds
SU(2) and SO(3). It can thus serve to detect spinorial sign changes even classically in a manner similar to how they are
detected quantum mechanically.
Recall that---as Aharonov and Susskind noted long ago---even quantum mechanically spinorial sign changes
cannot be detected in a direct or absolute manner \cite{Aharonov}. They showed that if fermions are shared coherently
between spatially isolated systems, then {\it relative} rotations of ${2\pi}$ between them may be observable
\cite{Hartung}\cite{Werner}\cite{Weingard}. Similarly,
in the above experiment two macroscopic bomb fragments would be rotating in tandem, {\it relative} to each other. Consequently,
the observables ${{sign}\,(\,+\,{\bf s}^k\cdot{\bf a})}$ and ${{sign}\,(\,-\,{\bf s}^k\cdot{\bf b})}$ would be sensitive to
the {\it relative} spinorial sign changes between these fragments. The correlation function ${{\cal E}({\bf a},\,{\bf b})}$
would then provide a measure of such sign changes in terms of geodesic distances on the manifolds SU(2) and SO(3).

It is also worth noting here that the observability of the spinorial sign changes in this manner is intimately tied up with
the fact that the quaternionic 3-sphere we defined in Eq.${\,}$(\ref{nonoonpara}) is naturally parallelized. In other words,
the metric on SU(2) is Euclidean, ${g_{\mu\nu}=\delta_{\mu\nu}}$, and hence the geodesic distances within it are dictated by
the function ${-\cos\eta_{{\bf a}{\bf b}}}$, as we noted in Eq.${\,}$(\ref{dismes}) and depicted in Fig.${\,}$\ref{fig-55}.
By contrast, correlation between the points of ${S^3}$ with a round metric and zero torsion cannot be stronger than the
linear correlation between the points of SO(3) \cite{Christian}. Thus it is
the discipline of parallelization in the manifold ${S^3\sim{\rm SU(2)}}$ which dictates that the correlation between
${{sign}\,(\,+\,{\bf s}^k\cdot{\bf a})}$ and ${{sign}\,(\,-\,{\bf s}^k\cdot{\bf b})}$ will be as strong as
${-\cos\eta_{{\bf a}{\bf b}}}$. Fortunately, parallelization happens to be the  natural property of the manifold
${S^3\sim{\rm SU(2)}}$, and hence the correlation function (\ref{correlations}) would provide us a natural means to detect
the signature of spinorial sign changes.  

One may wonder why such a non-trivial experiment is necessary when we can infer the topological properties of
${S^3\sim{\rm SU(2)}}$ by means of a simple classical device like Dirac's belt \cite{Hartung}. It should be noted, however,
that what the belt trick provides is only an indirect indication---by means of an external, {\it material} connection
(namely, the belt, or a cord)---of the rotational symmetry of SU(2). What it does not provide is a {\it direct} physical
evidence that the symmetry as non-trivial as SU(2) is indeed respected by our physical space. To establish that SU(2) is
indeed the symmetry group respected by our physical space even classically, what is needed is a demonstration of
spinorial sign changes between rotating objects even in the absence of any form of material interaction between them.

The experiment considered above would allows us to accomplish just such a ``belt-free'' demonstration. Once the bomb has
exploded and the two fragments are on their way towards the observation stations, there would be no material object (such
as a belt) connecting them. That is to say, once they have separated the stress-energy tensor between them would be zero
everywhere. The only physical link between them would be the orientation of the manifold ${S^3}$, which would provide a local
standard of reference for their mutual rotations and spinorial sign changes \cite{Weingard}.
In other words, only the orientation ${\lambda}$
of the 3-sphere would act as a non-material belt connecting the two fragments. In the following section we formalize
this ``belt without belt'' scenario in a systematic manner.  

\section{Distinguishing ${{\mathscr D}({\bf a},\,{\bf b})}$ on ${\rm SU(2)}$ from
${{\mathscr D}({\bf a},\,{\bf b})}$ on ${\rm SO(3)}$ by calculating ${{\cal E}({\bf a},\,{\bf b})}$}

The prescription for ${{\cal E}({\bf a},\,{\bf b})}$ above is how one would calculate the correlations in practice. Let us
now derive them theoretically, for the two cases considered in the sections II and III. To this end, we first note the
following definition of the orientation of a vector space \cite{Eberlein}\cite{Milnor}:

\begin{Joy}
An orientation of a finite dimensional vector space ${{\mathscr U}_d}$ is an equivalence class of
ordered basis, say ${\{f_1,\,\dots,\,f_d\}}$, which determines the same orientation of ${{\mathscr U}_d}$ as the basis
${\{f'_1,\,\dots,\,f'_d\}}$ if ${f'_i =  \omega_{ij} f_j}$ holds with ${det(\omega_{ij})>0}$, and the opposite orientation
of ${{\mathscr U}_d}$ as the basis ${\{f'_1,\,\dots,\,f'_d\}}$ if ${f'_i = \omega_{ij} f_j}$ holds with ${det(\omega_{ij}) < 0}$.
\end{Joy}

Thus each positive dimensional real vector space has precisely two possible orientations, which we shall
denote as ${\lambda=+1}$ or ${\lambda=-1}$. More generally an oriented smooth manifold such as ${S^3}$
consists of that manifold together with a choice of orientation for each of its tangent spaces.

It is important to note that orientation of a manifold is a {\it relative} concept \cite{Milnor}.
In particular, the orientation of a tangent space ${{\mathscr U}_d}$ of a manifold defined by the equivalence class of
ordered basis such as ${\{f_1,\,\dots,\,f_d\}}$ is meaningful only with respect to that defined by the equivalence
class of ordered basis ${\{f'_1,\,\dots,\,f'_d\}}$, and vice versa. To be sure, we can certainly orient a manifold
absolutely by choosing a set of ordered bases for all of its tangent spaces, but the resulting manifold can be said
to be left or right oriented only with respect of another such set of ordered basis.

Now the natural configuration space for the experiment considered above is a unit parallelized 3-sphere, which can
be embedded in ${{\rm I\!R}^4}$ with a choice of orientation, say ${\lambda=\pm\,1}$. This choice of orientation can be
identified with the initial state of the bomb fragments with respect to the orientation of the detector basis as follows.
We first characterize the embedding space ${{\rm I\!R}^4}$ by the graded basis
\begin{equation}
\left\{\,1,\;L_{1}(\lambda),\;L_{2}(\lambda),\;L_{3}(\lambda)\,\right\}, \label{sinbas}
\end{equation}
with ${\lambda = \pm\,1}$ representing the two possible orientations of ${S^3}$ and the basis elements
${L_{\mu}(\lambda)}$ satisfying the algebra
\begin{equation}
L_{\mu}(\lambda)\,L_{\nu}(\lambda) \,=\,-\,g_{\mu\nu}\,-\,\epsilon_{\mu\nu\rho}\,L_{\rho}(\lambda)\,. \label{wh-o88}
\end{equation}
As the notation suggests, we shall take the unit bivectors ${\{\,a_{\mu}\;L_{\mu}(\lambda)\,\}}$ to represent the spin angular
momenta of the bomb fragments. These momenta can then be assumed to be detected by the detector bivectors, say
${\{\,a_{\mu}\,D_{\mu}\,\}}$, with
the corresponding detector basis ${\left\{\,1,\,D_1,\,D_2,\,D_3\,\right\}}$ satisfying the algebra
\begin{equation}
D_{\mu}\,D_{\nu} \,=\,-\,g_{{\mu}{\nu}}\,-\,\epsilon_{{\mu}{\nu}{\rho}}\,D_{\rho}\, \label{wh-o99}
\end{equation}
and related to the spin basis ${\left\{\,1,\,L_{1}(\lambda),\,L_{2}(\lambda),\,L_{3}(\lambda)\,\right\}\;}$as
\begin{equation}
\left(\begin{array}{c} 1 \\ L_1(\lambda) \\ L_2(\lambda) \\ L_3(\lambda) \end{array}\right)\;=\;
\left(\begin{matrix} \; 1 \; & \; 0 \; & \; 0 \; & \; 0 \; \\
                     \; 0 \; & \; \lambda \; & \; 0 \; & \; 0 \; \\
                     \; 0 \; & \; 0 \; & \; \lambda \; & \; 0 \; \\
                     \; 0 \; & \; 0 \; & \; 0 \; & \; \lambda \;
\end{matrix}\right)
\left(\begin{array}{c} 1 \\ D_1 \\ D_2 \\ D_3 \end{array}\right)\!.
\end{equation}
Evidently, the determinant of this matrix works out to be ${det(\omega_{ij})= \lambda}$.
Since ${\lambda^2=+1}$ and ${\omega^2}$ is a ${4\times 4}$ identity matrix, this relation can be more succinctly written as
\begin{equation}
L_{\mu}(\lambda)\,=\,\lambda\,D_{\mu}\;\;\;\;\text{and}\;\;\;\;
D_{\mu}\,=\,\lambda\,L_{\mu}(\lambda)\,,\label{1237} 
\end{equation}
or equivalently as
\begin{equation}
\left\{\,1,\;L_{1}(\lambda),\;L_{2}(\lambda),\;L_{3}(\lambda)\,\right\}\,=\,
\left\{1,\,\lambda\,D_1,\,\lambda\,D_2,\,\lambda\,D_3\right\} \label{389-2}
\end{equation}
and
\begin{equation}
\left\{\,1,\;D_{1},\;D_{2},\;D_{3}\,\right\}\,=\,
\left\{1,\,\lambda\,L_1(\lambda),\,\lambda\,L_2(\lambda),\,\lambda\,L_3(\lambda)\right\}. \label{389-1}
\end{equation}
These relations reiterate the fact that orientation of any manifold is a {\it relative} concept. In particular, orientation of
${S^3}$ defined by the spin basis ${\{\,1,\,L_{\mu}(\lambda)\,\}}$ is meaningful only with respect to that defined by the detector
basis ${\{\,1,\,D_{\mu}\,\}}$ with the orientation ${\lambda=+1}$, and vice versa. Thus the spin basis are said to define the
{\it same} orientation of ${S^3}$ as the detector basis if ${L_{\mu}(\lambda=+1)=+D_{\mu}}$,
and the spin basis are said to define the {\it opposite}
orientation of ${S^3}$ as the detector basis if ${L_{\mu}(\lambda=-1)=-\,D_{\mu}}$.

We are now in a position to identify the
formal counterparts of the measurement variables ${{sign}\,({\bf s}^k\cdot{\bf a})=\pm\,1}$ and
${{sign}\,(\,-\,{\bf s}^k\cdot{\bf b})=\pm\,1}$ defined in the previous section:
\begin{align}
{sign}\,(\,+\,{\bf s}^k\cdot{\bf a})\,\equiv\,
{\mathscr A}({\bf a},\,{\lambda^k})\,&=\,\{-\,a_{\mu}\;D_{\mu}\,\}\,\{\,a_{\nu}\;L_{\nu}(\lambda^k)\,\} \notag \\
&=\,                      
\begin{cases}
+\,1\;\;\;\;\;{\rm if} &\lambda^k\,=\,+\,1 \\
-\,1\;\;\;\;\;{\rm if} &\lambda^k\,=\,-\,1
\end{cases} \label{88-oi}
\end{align}
and
\begin{align}
{sign}\,(\,-\,{\bf s}^k\cdot{\bf b})\,\equiv\,
{\mathscr B}({\bf b},\,{\lambda^k})\,&=\,\{+\,b_{\mu}\;D_{\mu}\,\}\,\{\,b_{\nu}\;L_{\nu}(\lambda^k)\,\} \notag \\
&=\,                      
\begin{cases}
-\,1\;\;\;\;\;{\rm if} &\lambda^k\,=\,+\,1 \\
+\,1\;\;\;\;\;{\rm if} &\lambda^k\,=\,-\,1\,,
\end{cases} \label{99-oi}
\end{align}
where the relative orientation ${\lambda}$ is now assumed to be a random variable, with 50/50 chance of
being ${+1}$ or ${-\,1}$ at the moment of the bomb-explosion considered in the previous section. We shall assume
that the orientation of ${S^3}$ defined by the detector basis ${\{\,1,\,D_{\nu}\,\}}$ has been fixed before hand \cite{Christian}. 
Thus the spin bivector ${\{\,a_{\mu}\;L_{\mu}(\lambda)\,\}}$ is a random bivector with its handedness determined
{\it relative} to the detector bivector ${\{\,a_{\nu}\;D_{\nu}\,\}}$, by the relation
\begin{equation}
{\bf L}({\bf a},\,\lambda)
\,\equiv\,\{\,a_{\mu}\;L_{\mu}(\lambda)\,\}\,=\,\lambda\,\{\,a_{\nu}\;D_{\nu}\,\}\,\equiv\,\lambda\,{\bf D}({\bf a}). \label{OJS}
\end{equation}
Using this relation the spin detection events (\ref{88-oi}) and (\ref{99-oi})
follow at once from the algebras (\ref{wh-o88}) and (\ref{wh-o99}).

It is important to note that the variables ${{\mathscr A}({\bf a},\,{\lambda})}$ and ${{\mathscr B}({\bf b},\,{\lambda})}$
given in equations (\ref{88-oi}) and (\ref{99-oi}) are generated with {\it different} bivectorial scales of dispersion (or
different standard deviations) for each measurement direction ${\bf a}$ and  ${\bf b}$. Consequently,
in statistical terms these variables are raw scores, as opposed to standard scores \cite{scores-1}.
Recall that a standard score, z, indicates how
many standard deviations an observation or datum is above or below the mean. If ${\rm x}$ is a raw (or unnormalized) score
and ${\overline{\rm x}}$ is its mean value, then the standard (or normalized) score, ${{\rm z}({\rm x})}$, is defined by
\begin{equation}
{\rm z}({\rm x})\,=\,\frac{{\rm x}\,-\,{\overline{\rm x}}}{\sigma({\rm x})}\,,
\end{equation}
where ${\sigma({\rm x})}$ is the standard deviation of ${\rm x}$. A standard score thus represents the distance between
a raw score and population mean in the units of standard deviation, and
allows us to make comparisons of raw scores that come from very different sources \cite{Christian}\cite{scores-1}.
In other words, the mean value of the standard score itself is always zero, with standard deviation unity.
In terms of these concepts the correlation between raw scores ${\rm x}$ and ${\rm y}$ is defined as
\begin{align}
{\cal E}({\rm x},\,{\rm y})\,&=\;\frac{\,{\displaystyle\lim_{\,n\,\gg\,1}}\left[{\displaystyle\frac{1}{n}}\,
{\displaystyle\sum_{k\,=\,1}^{n}}\,({\rm x}^k\,-\,{\overline{\rm x}}\,)\;
({\rm y}^k\,-\,{\overline{\rm y}}\,)\right]}{\sigma({\rm x})\;\sigma({\rm y})} \label{co} \\
&=\,\lim_{\,n\,\gg\,1}\left[\frac{1}{n}
\sum_{k\,=\,1}^{n}\,{\rm z}({\rm x}^k)\;{\rm z}({\rm y}^k)\right]. \label{stan}
\end{align}
It is vital to appreciate that covariance by itself---{\it i.e.}, the numerator of Eq.${\,}$(\ref{co}) by itself---does not
provide the correct measure of association between the raw scores, not the least because it depends on different units and scales
(or different scales of dispersion) that may have been used in the measurements of such scores. Therefore, to arrive at the correct
measure of association between the raw scores one must either use Eq.${\,}$(\ref{co}), with the product of standard
deviations in the denominator, or use covariance of the standardized variables, as in Eq.${\,}$(\ref{stan}).

Now, as noted above, the random variables ${{\mathscr A}({\bf a},\,{\lambda})}$ and ${{\mathscr B}({\bf b},\,{\lambda})}$ are
products of two factors---one random and another non-random. Within the variable ${{\mathscr A}({\bf a},\,{\lambda})}$
the bivector ${{\bf L}({\bf a},\,\lambda)}$ is a random factor---a function of the orientation ${\lambda}$, whereas
the bivector ${-\,{\bf D}({\bf a})}$ is a non-random factor, independent of the orientation ${\lambda}$. Thus,
as random variables, ${{\mathscr A}({\bf a},\,{\lambda})}$ and ${{\mathscr B}({\bf b},\,{\lambda})}$ are
generated with {\it different} standard deviations---{\it i.e.}, {\it different} sizes of the typical error.
More specifically, ${{\mathscr A}({\bf a},\,{\lambda})}$ is generated with the standard deviation
${-\,{\bf D}({\bf a})}$, whereas ${{\mathscr B}({\bf b},\,{\lambda})}$ is generated
with the standard deviation ${+\,{\bf D}({\bf b})}$. These two deviations
can be calculated as follows. Since errors in the linear relations propagate linearly,
the standard deviation ${\sigma({\mathscr A}\,)}$ of ${{\mathscr A}({\bf a},\,{\lambda})}$ is equal to ${-\,{\bf D}({\bf a})}$
times the standard deviation of ${{\bf L}({\bf a},\,\lambda)}$ [which we shall denote as
${\sigma({A})=\sigma({\bf L}_{\bf a})}$], whereas the standard deviation ${\sigma({\mathscr B}\,)}$ of
${{\mathscr B}({\bf b},\,{\lambda})}$ is equal to ${+\,{\bf D}({\bf b})}$ times the standard deviation of
${{\bf L}({\bf b},\,\lambda)}$ [which we shall denote as ${\sigma({B})=\sigma({\bf L}_{\bf b})}$]:
\begin{align}
\sigma({\mathscr A}\,)\,&=\,-\,{\bf D}({\bf a})\,\sigma({A}) \notag \\
\text{and}\;\;\sigma({\mathscr B}\,)\,&=\,+\,{\bf D}({\bf b})\,\sigma({B}).
\end{align}
But since the bivector ${{\bf L}({\bf a},\,\lambda)}$ is normalized to unity, and since its mean
value ${m({\bf L}_{\bf a})}$ vanishes on the account of ${\lambda}$ being a
fair coin, its standard deviation is easy to calculate, and it turns out to be equal to unity:
\begin{align}
\sigma({A})\,&=\,\sqrt{\frac{1}{n}\sum_{k\,=\,1}^{n}\,\left|\left|\,A({\bf a},\,{\lambda}^k)\,-\,
{\overline{A({\bf a},\,{\lambda}^k)}}\;\right|\right|^2\,}\, \notag \\
&=\,\sqrt{\frac{1}{n}\sum_{k\,=\,1}^{n}\,
\left|\left|\,{\bf L}({\bf a},\,\lambda^k)\,-\,0\,\right|\right|^2\,}\,=\,1, \label{var-c}
\end{align}
where the last equality follows from the normalization of ${{\bf L}({\bf a},\,\lambda)}$.
Similarly, we find that ${\sigma({B})}$ is also equal to ${1}$. Consequently, the standard deviation of
${{\mathscr A}({\bf a},\,{\lambda})=\pm\,1}$ works out to be ${-\,{\bf D}({\bf a})}$, and the standard
deviation of ${{\mathscr B}({\bf b},\,{\lambda})=\pm\,1}$ works out to be ${+\,{\bf D}({\bf b})}$. Putting
these two results together, we arrive at the following standard scores corresponding to the raw scores
${\mathscr A}$ and ${\mathscr B}$:
\begin{align}
A({\bf a},\,{\lambda})&=\frac{\,{\mathscr A}({\bf a},\,{\lambda})\,-\,
{\overline{{\mathscr A}({\bf a},\,{\lambda})}}}{\sigma({\mathscr A})} \notag \\
\,&=\,\frac{\,-\,{\bf D}({\bf a})\,{\bf L}({\bf a},\,\lambda)\,-\,0\,}{-\,{\bf D}({\bf a})}
\,=\,{\bf L}({\bf a},\,\lambda) \label{var-a}
\end{align}
and
\begin{align}
B({\bf b},\,{\lambda})&=\frac{\,{\mathscr B}({\bf b},\,{\lambda})\,-\,
{\overline{{\mathscr B}({\bf b},\,{\lambda})}}}{\sigma({\mathscr B})} \notag \\
\,&=\,\frac{\,+\,{\bf D}({\bf b})\,{\bf L}({\bf b},\,\lambda)\,-\,0\,}{+\,{\bf D}({\bf b})}
\,=\,{\bf L}({\bf b},\,\lambda),\label{var-b}
\end{align}
where we have used identities such as ${-\,{\bf D}({\bf a}){\bf D}({\bf a})=+1}$. In the Appendix B below we shall derive the
results (\ref{var-c}), (\ref{var-a}), and (\ref{var-b}) in a much more systematic manner.

Now, since we have assumed that initially there was 50/50 chance between the right-handed and left-handed orientations
of the 3-sphere ({\it i.e.}, equal chance between the initial states ${{\lambda}=+\,1}$ and ${{\lambda}=-\,1}$), the
expectation values of the raw scores ${{\mathscr A}({\bf a},\,{\lambda})}$ and ${{\mathscr B}({\bf b},\,{\lambda})}$
vanish identically. On the other hand, as discussed above, the correlation between these raw scores can
be obtained only by computing the covariance between the corresponding standardized variables
${{A}({\bf a},\,{\lambda})}$ and ${{B}({\bf b},\,{\lambda})}$, which gives
\begin{align}
{\cal E}&({\bf a},\,{\bf b})\,
=\,\lim_{\,n\,\gg\,1}\left[\frac{1}{n}\sum_{k\,=\,1}^{n}\,A({\bf a},\,{\lambda}^k)\,B({\bf b},\,{\lambda}^k)\right] \notag \\
&\;\;\;=\lim_{\,n\,\gg\,1}\left[\frac{1}{n}\sum_{k\,=\,1}^{n}\,
\left\{\,a_{\mu}\;L_{\mu}(\lambda^k)\,\right\}\,\left\{\,b_{\nu}\;L_{\nu}(\lambda^k)\,\right\}\right] \notag \\
&\;\;\;=\,-\,g_{\mu\nu}\,a_{\mu}\,b_{\nu}\,-\lim_{\,n\,\gg\,1}\left[\frac{1}{n}\sum_{k\,=\,1}^{n}\,
\left\{\,\epsilon_{\mu\nu\rho}\;a_{\mu}\,b_{\nu}\;L_{\rho}(\lambda^k)\,\right\}\right] \notag \\
&\;\;\;=\,-\,g_{\mu\nu}\,a_{\mu}\,b_{\nu}\,-\lim_{\,n\,\gg\,1}\left[\frac{1}{n}\sum_{k\,=\,1}^{n}\,
\lambda^k\right]\left\{\,\epsilon_{\mu\nu\rho}\;a_{\mu}\,b_{\nu}\;D_{\rho}\,\right\} \notag \\
&\;\;\;=\,-\,g_{\mu\nu}\,a_{\mu}\,b_{\nu}\,-\,0\,, \label{corwedid}
\end{align}
where we have used the algebra (\ref{wh-o88}) and relation (\ref{OJS}).
Consequently, as explained in the paragraph just below Eq.${\,}$(\ref{stan}),
when the raw scores ${{\mathscr A}=\pm\,1}$ and ${{\mathscr B}=\pm\,1}$ are
compared, their product moment will inevitably yield
\begin{align}
{\cal E}({\bf a},\,{\bf b})\,&=\lim_{\,n\,\gg\,1}\left[\frac{1}{n}\sum_{k\,=\,1}^{n}\,
{\mathscr A}({\bf a},\,{\lambda}^k)\;{\mathscr B}({\bf b},\,{\lambda}^k)\right]\, \notag \\
&=\,-\,g_{\mu\nu}\,a_{\mu}\,b_{\nu}\,, \label{darcor}
\end{align}
since the correlation between the raw scores ${{\mathscr A}}$ and ${\mathscr B}$ is
equal to covariance between the standard scores ${A}$ and${\;B}$.
[See also Appendix C below for the upper bound on ${\cal E}$.]

So far in this section we have put no restrictions on the metric tensor, which, in the normal coordinates centered at ${p}$
would be of the form
\begin{equation}
g_{\mu\nu}(x)\,=\,\delta_{\mu\nu}\,-\,\frac{1}{3}\;
{\mathscr R}_{\,\alpha\,\mu\,\nu\,\gamma}\,x^{\alpha}\,x^{\gamma}\,+\,O\left(|x|^3\right).
\end{equation}
In other words, the algebra (\ref{wh-o88}) we have used in the derivation of the correlation (\ref{darcor})
is a general Clifford algebra, with no restrictions placed on the quadratic form
${\langle\,\cdot\;,\,\cdot\,\rangle}$ \cite{Frankel}. On the other hand, in the previous
sections we saw that there are at least two possibilities for the metric tensor ${g_{\mu\nu}}$
corresponding to the two geometries of the rotation groups SU(2) and SO(3) (cf. Fig.${\,}$\ref{fig-55}):
\begin{equation}
-\,g_{\mu\nu}\,a_{\mu}\,b_{\nu}\,=\,-\,\delta_{\mu\nu}\,a_{\mu}\,b_{\nu}\,=\,-\,\cos\eta_{{\bf a}{\bf b}}\,, \label{hjro}
\end{equation}
which manifests sensitivity to spinorial sign changes, and
\begin{align}
-\,g_{\mu\nu}\,a_{\mu}\,b_{\nu}\,&=\,-\,J_{\mu\nu}\,a_{\mu}\,b_{\nu}\,=\,-\,\cos\alpha_{{\bf a}{\bf b}} \notag \\
&=\,
\begin{cases}
-\,1\,+\,\frac{2}{\pi}\,\eta_{{\bf a}{\bf b}}
\;\;\;{\rm if} &\!\! 0 \leq \eta_{{\bf a}{\bf b}} \leq \pi \\
\\
+\,3\,-\,\frac{2}{\pi}\,\eta_{{\bf a}{\bf b}}
\;\;\;{\rm if} &\!\! \pi \leq \eta_{{\bf a}{\bf b}} \leq 2\pi\,, \label{equawhichr-2}
\end{cases}
\end{align}
which manifests insensitivity to spinorial sign changes. These two possibilities amount to identifying the spin basis
${\{\,1,\,L_{\mu}(\lambda)\,\}}$ either with the basis ${\{\,1,\,\beta_{\mu}(\lambda)\,\}}$ of ${T_{\bf q}S^3}$ defined
in Eq.${\,}$(\ref{orig-alg}) or with the basis ${\{\,1,\,\xi_{\mu}(\lambda)\,\}}$ of ${T_{[{\bf q}]}{\rm I\!R}{\rm P}^3}$
defined in Eq.${\,}$(\ref{norig-alg}). The two metrics ${\delta_{\mu\nu}}$ and ${J_{\mu\nu}}$, respectively, are therefore
measures of the geodesic distances on the manifolds ${S^3}$ and ${{\rm I\!R}{\rm P}^3}$, as we discussed in the previous
sections. Thus, the correlation ${{\cal E}({\bf a},\,{\bf b})}$ we derived in Eq.${\,}$(\ref{corwedid}) can
serve to distinguish the geodesic distances ${{\mathscr D}({\bf a},\,{\bf b})}$ on the groups SU(2) and SO(3). 

\section{Conclusion}

In their landmark textbook on gravitation Misner, Thorne, and Wheeler noted that there is something about the geometry of
orientation that is not fully taken into account in the usual concept of orientation \cite{Misner}\cite{Penrose}\cite{Hartung}.
They noted that rotations in space by ${0}$, ${\pm\,4\pi}$, ${\pm\,8\pi,\,\dots}$, leave all objects in their standard
orientation-entanglement relation with their surroundings, whereas rotations by ${\pm\,2\pi}$, ${\pm\,6\pi}$,
${\pm\,10\pi,\,\dots}$, restore only their orientation but not their orientation-entanglement relation with their surroundings.
The authors wondered whether there was a detectable difference in physics
for the two inequivalent states of an object. Earlier Aharonov and Susskind had argued that there
{\it is} a detectable difference for
such states in quantum physics, but not classical physics, where both
absolute and relative ${2\pi}$ rotations are undetectable \cite{Aharonov}.

In this paper we have argued that there is, in fact, a detectable difference between absolute and relative ${2\pi}$ rotations
even in classical physics. In particular, we have demonstrated the observability of spinorial sign changes under ${2\pi}$
rotations in terms of geodesic distances on the group manifolds SU(2) and SO(3). Moreover, we have proposed a
macroscopic experiment which could infer the ${4\pi}$ periodicity in principle \cite{Aharonov}. The proposed experiment
has the potential to transform our understanding of the relationship between classical and quantum physics \cite{Christian}.

\vspace{-0.15cm}

\acknowledgments
I am grateful to Lucien Hardy for several months of correspondence which led to significant improvements
in the contents and presentation of the Appendix B below. I am also grateful to Martin Castell for his kind
hospitality in the Materials Department of the University of Oxford, and to Manfried Faber and Christian Els
for comments on the earlier versions of this paper. Christian Els also kindly carried out parts of the
calculation in Appendix A, especially the derivation of the curvature in Eq.${\,}$(\ref{vancur}). 
This work was funded by a grant from the Foundational Questions Institute (FQXi) Fund, a donor advised fund
of the Silicon Valley Community Foundation on the basis of proposal FQXi-MGA-1215 to the Foundational Questions
Institute. I thank Jurgen Theiss of Theiss Research (La Jolla, CA) for administering the grant on my behalf.

\appendix
\section{Parallelizing Torsion in ${S^3}$ and ${{\rm I\!R}{\rm P}^3}$}

In this appendix we show that both ${{\rm I\!R}{\rm P}^3\sim{\rm SO(3)}}$ and its covering space ${S^3\sim{\rm SU(2)}}$
can be characterized by torsion alone, since their Riemann curvatures vanish identically with respect to the Weitzenb\"ock
connection. To this end, we begin by defining possible basis vectors in ${{\rm I\!R}^4}$, which we have taken to be the
embedding space:
\begin{equation}
\left\{\begin{matrix} \, +{q}_0 \, & \, +{q}_1 \, & \, +{q}_2 \, & \, +{q}_3 \, \\
                              \, -{q}_1 \, & \, +{q}_0 \, & \, +{q}_3 \, & \, -{q}_2 \, \\
                               \, -{q}_2 \, & \, -{q}_3 \, & \, +{q}_0 \, & \, +{q}_1 \, \\
                                \, -{q}_3 \, & \, +{q}_2 \, & \, -{q}_1 \, & \, +{q}_0 \,
\end{matrix}\right\}\!. \label{array0}
\end{equation}
Either the four column vectors or the four row vectors of this array may be taken to form an orthonormal basis of ${{\rm I\!R}^4}$.
If we take the first row of the array to represent the points of ${S^3}$ as in definition (\ref{nonoonpara}), then, as we derived
in Eqs.${\,}$(\ref{inthederi-1}) to (\ref{inthederi-3}), the remaining rows of the array provide the basis of the tangent space
${T_{\bf q}S^3}$ at each point of ${S^3}$:
\begin{align}
\beta_1({\bf q})\,=\,({{\bf e}_2}\,\wedge\,{{\bf e}_3})\,{\bf q}
\,&=\,(-q_1,\;+q_0,\;+q_3,\;-q_2), \label{inthederi-133} \\
\beta_2({\bf q})\,=\,({{\bf e}_3}\,\wedge\,{{\bf e}_1})\,{\bf q}
\,&=\,(-q_2,\;-q_3,\;+q_0,\;+q_1), \label{inthederi-233} \\
\beta_3({\bf q})\,=\,({{\bf e}_1}\,\wedge\,{{\bf e}_2})\,{\bf q}
\,&=\,(-q_3,\;+q_2,\;-q_1,\;+q_0). \label{inthederi-333}
\end{align}
Similarly, in a dual description, if we take the leftmost column of the array
(\ref{array0}) to represent the points of ${S^3}$, then the remaining columns
of the array provide the basis of the cotangent space ${T^*_{\bf q}S^3}$ at each point of ${S^3}$:
\begin{align}
\beta^1({\bf q}^{\dagger})\,=\,({{\bf e}_2}\,\wedge\,{{\bf e}_3})\,{\bf q}^{\dagger}
\,&=\,(+q_1,\;+q_0,\;-q_3,\;+q_2), \label{inthederi-13377} \\
\beta^2({\bf q}^{\dagger})\,=\,({{\bf e}_3}\,\wedge\,{{\bf e}_1})\,{\bf q}^{\dagger}
\,&=\,(+q_2,\;+q_3,\;+q_0,\;-q_1), \label{inthederi-23388} \\
\beta^3({\bf q}^{\dagger})\,=\,({{\bf e}_1}\,\wedge\,{{\bf e}_2})\,{\bf q}^{\dagger}
\,&=\,(+q_3,\;-q_2,\;+q_1,\;+q_0). \label{inthederi-33399}
\end{align}
For our calculations we shall use the index notation ${\beta^i_{\mu}}$ for the basis elements and
${\beta_i^{\mu}}$ for their inverses, with the index ${\mu=1,\,2,\,3}$ representing the tangent space axes
and the index ${i=0,\,1,\,2,\,3}$ representing the embedding space axes. 
Then the matrices ${\beta^i_{\mu}}$ and their inverses ${\beta_i^{\mu}}$ satisfy
\begin{equation}
\beta_i^{\mu}\,\beta^i_{\nu}\,=\,\delta_{\;\nu}^{\mu}\;\;\;\text{and}\;\;\;
\beta^i_{\mu}\,\beta_j^{\mu}\,=\,\delta_{\,j}^{i}\,, \label{no-orth}
\end{equation}
with ${\delta_{\;\nu}^{\mu}}$ as a ${3\times 3}$ sub-matrix and ${\delta_{\,j}^{i}}$ as a ${4\times 4}$ matrix.

Now in a parallelized 3-sphere a spinor ${{\bf v}({\bf p})\in T_{\bf p}S^3}$ is said to be absolutely parallel to a spinor
${{\bf w}({\bf q})\in T_{\bf q}S^3}$ if all of the {\it components} of ${{\bf v}({\bf p})}$ at ${T_{\bf p}S^3}$ are equal
to those of ${{\bf w}({\bf q})}$ at ${T_{\bf q}S^3}$. That is to say, if ${{\bf v}({\bf p})=v^{\mu}\beta_{\mu}({\bf p})}$
and ${{\bf w}({\bf q})=w^{\mu}\beta_{\mu}({\bf q})}$, then ${v^{\mu}=w^{\mu}}$. For a parallelized 3-sphere we therefore
require that the components of
any spinor ${\bf v}({\bf p})$ at a point ${p\in S^3}$ remain the same when it is parallel transported to a nearby point
${p+\epsilon\in S^3}$ \cite{telepar}:
\begin{equation}
v^{\mu}(p)\beta^i_{\mu}(p)=v^{\mu}(p+\epsilon)\beta^i_{\mu}(p+\epsilon). \label{rellll}
\end{equation}
By expanding the right hand side up to terms of order $\epsilon$ we obtain
\begin{align}
v^{\mu}&(p+\epsilon)\beta^i_{\mu}(p+\epsilon) \notag \\
&=\left[v^{\mu}(p)-\epsilon^{\nu}v^{\alpha}(p)\Omega_{\nu\alpha}^{\mu}(p)\right]\left[\beta^i_{\mu}(p)+
\epsilon^{\nu}\partial_{\nu}\beta^i_{\mu}(p)\right] \notag \\
&=v^{\mu}(p)\beta^i_{\mu}(p) \notag \\
&\;\;\;\;-\epsilon^{\nu}v^{\alpha}(p)
\left[\Omega_{\nu\alpha}^{\mu}(p)\beta^i_{\mu}(p)-\partial_{\nu}\beta^i_{\alpha}(p)\right]+O(\epsilon^2)\,,
\end{align}
where ${\Omega_{\nu\alpha}^{\mu}}$ are the connection coefficients. Evidently, the second term of this equation must vanish
for the relation (\ref{rellll}) to hold. This gives the connection coefficients ${\Omega_{\nu\alpha}^{\mu}}$ in terms of the
partial derivatives of the basis elements:
\begin{align}
\Omega_{\nu\alpha}^{\mu}(p)\beta^i_{\mu}(p)=\partial_{\nu}\beta^i_{\alpha}(p).
\end{align}
Contracting this relation with the basis elements ${\beta_i^{\sigma}(p)}$ then leads to the {\it Weitzenb\"ock connection}
[cf. Eq.${\,}$(\ref{W-con})]:
\begin{align}
\Omega_{\nu\alpha}^{\mu}(p)=\beta_i^{\mu}(p)\partial_{\nu}\beta^i_{\alpha}(p)\,. \label{asion}
\end{align}
Alternatively (but equivalently), absolute parallelism on ${S^3}$ can be defined by requiring that the basis elements
${\beta^i_{\mu}}$ remain covariantly constant during parallel transport:
\begin{equation}
\nabla_{\alpha}\beta^i_{\nu}:=\,\partial_{\alpha}\beta^i_{\nu}-\,\Omega_{\alpha\nu}^{\mu}\beta^i_{\mu}=\,0\,.
\end{equation}
Solving this equation for ${\Omega^{\mu}_{\nu\alpha}}$ again gives the connection obtained in (\ref{asion}).
We can now evaluate the curvature tensor of ${S^3}$ with respect to this asymmetric connection:
\begin{align}
{\mathscr R}^{\sigma}_{\;\alpha\mu\nu}&[\,S^3\,]
=\partial_{\mu}\Omega^{\sigma}_{\nu\alpha}-\partial_{\nu}\Omega^{\sigma}_{\mu\alpha}+
\Omega^{\lambda}_{\nu\alpha}\Omega^{\sigma}_{\mu\lambda} - \Omega^{\lambda}_{\mu\alpha}\Omega^{\sigma}_{\nu\lambda} \notag \\
&\!\!=\partial_{\mu}(\beta_i^{\sigma}\partial_{\nu}\beta^i_{\alpha})-\partial_{\nu}(\beta_i^{\sigma}
\partial_{\mu}\beta^i_{\alpha}) \notag \\
&\!\!\;\;\;\;\;\;\;\;\;\;\;\;\;\;\;+\beta_k^{\sigma}\partial_{\mu}\beta^k_{\lambda}\,\beta_i^{\lambda}\partial_{\nu}\beta^i_{\alpha}
-\beta_k^{\sigma}\partial_{\nu}\beta^k_{\lambda}\,\beta_i^{\lambda}\partial_{\mu}\beta^i_{\alpha} \notag \\
&\!\!\!=\partial_{\mu}\beta_i^{\sigma}\partial_{\nu}\beta^i_{\alpha}+\beta_i^{\sigma}\partial_{\mu}\partial_{\nu}\beta^i_{\alpha}
-\partial_{\nu}\beta_i^{\sigma}\partial_{\mu}\beta^i_{\alpha}-\beta_i^{\sigma}\partial_{\nu}\partial_{\mu}\beta^i_{\alpha} \notag \\
&\!\!\;\;\;\;\;\;\;\;\;\;\;\;\;\;\;-\beta_i^{\sigma}\beta^i_{\lambda}\partial_{\mu}\beta_k^{\lambda}\partial_{\nu}\beta^k_{\alpha}+
\beta_i^{\sigma}\beta^i_{\lambda}\partial_{\nu}\beta_k^{\lambda}\partial_{\mu}\beta^k_{\alpha} \notag \\
&\!\!\!=\partial_{\mu}\beta_i^{\sigma}\partial_{\nu}\beta^i_{\alpha}-\partial_{\nu}\beta_i^{\sigma}\partial_{\mu}\beta^i_{\alpha}
-\partial_{\mu}\beta_i^{\sigma}\partial_{\nu}\beta^i_{\alpha}+\partial_{\nu}\beta_i^{\sigma}\partial_{\mu}\beta^i_{\alpha} \notag \\
&\!\!\!=0\,. \label{vancur}
\end{align}
Thus the curvature of ${S^3}$ with respect to Weitzenb\"ock connection vanishes identically. The geometric
properties of the quaternionic 3-sphere are thus entirely captured by the parallelizing torsion,
which is evaluated in Eq.${\,}$(\ref{torsioninthe}). In the present index notation it can be expressed as
\begin{equation}
{\mathscr T}_{\,\mu\,\nu}^{\,\sigma}[\,S^3\,]=\Omega_{\mu\nu}^{\sigma}\,-\,\Omega_{\nu\mu}^{\sigma}=
\beta_i^{\sigma}\left(\partial_{\mu}\beta^i_{\nu}\,-\,\partial_{\nu}\beta^i_{\mu}\right). \label{tor1}
\end{equation}

It is important to recognize here that the quotient map ${\varphi:S^3\rightarrow{\rm I\!R}{\rm P}^3}$ we discussed in section
III to obtain ${{\rm I\!R}{\rm P}^3}$ from ${S^3}$ is a {\it local} isometry. The infinitesimal map
${d\varphi({\bf q}):\,T_{\bf q}S^3\rightarrow T_{[\,{\bf q}\,]}{\rm I\!R}{\rm P}^3}$
is therefore an isometry. The inner product in ${T_{\bf q}S^3}$ defined by (\ref{orige-flat}) at each point
of ${S^3}$ is thus preserved only locally under the action of ${\varphi}$:
\begin{align}
\langle\,\{a_{\mu}\,{\beta}_{\mu}({\bf q})&\},\;\{b_{\nu}\,{\beta}_{\nu}({\bf q})\}\,\rangle \notag \\
&=\,\langle\,\{a_{\mu}\,{\xi}_{\mu}[\varphi({\bf q})]\},\;\{b_{\nu}\,{\xi}_{\nu}[\varphi({\bf q})]\}\,\rangle\,,
\end{align}
where ${\{\,\xi_{\mu}[\varphi({\bf q})]\}}$ are the basis defining ${T_{[\,{\bf q}\,]}{\rm I\!R}{\rm P}^3}$.
Thus, because of the presence of torsion, the rule for parallel transporting a spinor from one point
to another on ${{\rm I\!R}{\rm P}^3}$ is not preserved by the map ${\varphi:S^3\rightarrow{\rm I\!R}{\rm P}^3}$.
It is given by a different Weitzenb\"ock connection, defined by ${\xi_{\mu}}$:
\begin{align}
\widehat{\Omega}_{\nu\alpha}^{\mu}(p)=\xi_i^{\mu}(p)\partial_{\nu}\xi^i_{\alpha}(p)\,.
\end{align}
Since the basis ${\{\,\xi_{\mu}\}}$ are covariantly constant with respect to ${\widehat{\Omega}_{\nu\alpha}^{\mu}}$,
the curvature of ${{\rm I\!R}{\rm P}^3}$ also vanishes identically:
\begin{equation}
{\mathscr R}^{\sigma}_{\;\alpha\mu\nu}[\,{\rm I\!R}{\rm P}^3\,]\,=\,0\,.
\end{equation}
Despite the fat that the metric tensor ${J_{\mu\nu}}$ on ${{\rm I\!R}{\rm P}^3}$ is no longer Euclidean
[cf. Eq.${\,}$(\ref{orige-onoflat})], the steps in the derivation analogous to (\ref{vancur}) go through because, just
as in (\ref{no-orth}), the matrices ${\xi^i_{\mu}}$ and their inverses ${\xi_i^{\mu}}$ continue to satisfy
\begin{equation}
\xi_i^{\mu}\,\xi^i_{\nu}\,=\,\delta_{\;\nu}^{\mu}\;\;\;\text{and}\;\;\;
\xi^i_{\mu}\,\xi_j^{\mu}\,=\,\delta_{\,j}^{i}\,. \label{conno-orth}
\end{equation}
These are simply reciprocal relations between matrices and their inverses and not the orthonormality relations
for the basis. Consequently, the manifold ${{\rm I\!R}{\rm P}^3}$ remains as parallelized as ${S^3}$ under
the map ${\varphi:S^3\rightarrow{\rm I\!R}{\rm P}^3}$. This state of affairs, in fact, forms the basis
of some powerful theorems in the mathematics of division algebras \cite{Christian}.

What {\it does} change under the map ${\varphi:S^3\rightarrow{\rm I\!R}{\rm P}^3}$ is the characteristic
torsion within the parallelized manifold:
\begin{equation}
{\mathscr T}_{\,\mu\,\nu}^{\,\sigma}[\,{\rm I\!R}{\rm P}^3\,]=\widehat{\Omega}_{\mu\nu}^{\sigma}
\,-\,\widehat{\Omega}_{\nu\mu}^{\sigma}=
\xi_i^{\sigma}\left(\partial_{\mu}\xi^i_{\nu}\,-\,\partial_{\nu}\xi^i_{\mu}\right). \label{tor2}
\end{equation}
The difference between the two expressions of the torsion, namely (\ref{tor1}) and (\ref{tor2}),
is seen more transparently in the notation of geometric algebra used in the derivation of
(\ref{torsioninthe}). In this notation the torsion within ${S^3}$ is given by
\begin{equation}
{\mathscr T}[{\boldsymbol\beta}({\bf a}),\,{\boldsymbol\beta}({\bf b})]\,=\,{\bf a}\,\wedge\,{\bf b}
\,=\,{\boldsymbol\beta}({\bf c})\,\sin\eta_{\bf ab}\,,\label{torsnot}
\end{equation}
whereas that within ${{\rm I\!R}{\rm P}^3}$ is given by
\begin{equation}
{\mathscr T}[{\boldsymbol\xi}({\bf a}),\,{\boldsymbol\xi}({\bf b})]\,=\,{\bf a}\,\wedge\,{\bf b}
\,=\,{\boldsymbol\xi}({\bf c})\,\sin\alpha_{\bf ab}\,.\label{nottors}
\end{equation}
Here ${{\bf c}={\bf a}\times{\bf b}/|{\bf a}\times{\bf b}|}$, and---as we discussed in section III---the
angles ${\eta_{\bf ab}}$ and ${\alpha_{\bf ab}}$ are non-linearly related as
\begin{equation}
\sin\alpha_{{\bf a}{\bf b}}\,=\,
\begin{cases}
+\,\frac{2}{\pi}\,\eta_{{\bf a}{\bf b}}
&\;{\rm if}\;-\frac{\pi}{2} \leq \eta_{{\bf a}{\bf b}} \leq \frac{\pi}{2} \\
\\
+\,2\,-\,\frac{2}{\pi}\,\eta_{{\bf a}{\bf b}}
&\;{\rm if}\;\;\;\;\;\frac{\pi}{2} \leq \eta_{{\bf a}{\bf b}} \leq \frac{3\pi}{2}\,. \label{whrrr}
\end{cases}
\end{equation}
But since the map ${\varphi:S^3\rightarrow{\rm I\!R}{\rm P}^3}$ is a local isometry, the bivectors ${{\boldsymbol\beta}({\bf c})}$
and ${{\boldsymbol\xi}({\bf c})}$ representing a binary rotation about ${\bf c}$ are the same. Consequently, the torsion within
${{\rm I\!R}{\rm P}^3}$ in terms of the Euclidean angle ${\eta_{\bf ab}}$ is given by
\begin{equation}
{\mathscr T}_{{\bf a}\,{\bf b}}\,=\,{\boldsymbol\beta}({\bf c})\times
\begin{cases}
+\,\frac{2}{\pi}\,\eta_{{\bf a}{\bf b}}
&\;{\rm if}\;-\frac{\pi}{2} \leq \eta_{{\bf a}{\bf b}} \leq \frac{\pi}{2} \\
\\
+\,2\,-\,\frac{2}{\pi}\,\eta_{{\bf a}{\bf b}}
&\;{\rm if}\;\;\;\;\;\frac{\pi}{2} \leq \eta_{{\bf a}{\bf b}} \leq \frac{3\pi}{2}\,. \label{whiiiiii}
\end{cases}
\end{equation}
Comparing this expression with (\ref{torsnot}) we now clearly see the difference between the parallelizing torsions within
${S^3}$ and ${{\rm I\!R}{\rm P}^3}$. It is this difference that is reflected in Fig.${\,}$\ref{fig-55}.

\section{Error Propagation in ${S^3}$}

In this appendix we spell out the statistical basis of the results (\ref{var-c}), (\ref{var-a}),
and (\ref{var-b}) in greater detail. To this end, let a probability density function
${P({\bf q}):S^3\rightarrow\,[{\hspace{1pt}}0,\,1]}$ of random quaternions over ${S^3}$ be defined as:
\begin{equation}
P({\bf q})\,=\,\frac{1}{\sqrt{2\pi\left|\left|{\hspace{1pt}}\sigma({\bf q})\right|\right|^2\,}\;}\,
\exp\left\{-\,\frac{\,\left|\left|\,{\bf q}-m({\bf q})\right|\right|^2}{2\,
\left|\left|{\hspace{1pt}}\sigma({\bf q})\right|\right|^2}\right\},\label{probdensi-8}
\end{equation}
where the square root of ${\bf q=p\,p}$, ${{\bf p}\in S^3}$, is defined as
\begin{equation}
\sqrt{\,{\bf q}\,}\,=\sqrt{\,{\bf p}\,{\bf p}\,}\,:=\,\pm\,{\bf p}^{\dagger}(\,{\bf p}\,{\bf p}\,)
\,=\,\pm(\,{\bf p}^{\dagger}{\bf p}\,)\,{\bf p}\,=\,\pm\,{\bf p}\,. \label{1root1}  
\end{equation}
It is a matter of indifference whether the distribution of ${{\bf q}\in S^3}$ so chosen happens to be normal or not. Here
\begin{equation}
{\bf q}^k(\psi,\,{\bf r},\,\lambda)\,:=\,\left\{\lambda^k\,\cos\frac{\psi}{2}\,+\,{\bf L}\!\left({\bf r},\,\lambda^k\right)
\,\sin\frac{\psi}{2}\,\right\} \label{quat-0}
\end{equation}
is an arbitrary quaternion within ${S^3(\lambda)}$ of the form
\begin{equation}
{\bf q}={\mathscr A}+\,{\bf L}=\text{scalar}+\text{bivector}\,,
\end{equation}
with ${0\leq\psi\leq 4\pi}$ being the double-covering rotation angle about ${\bf r}$-axis. The mean value of
${{\bf q}(\psi,\,{\bf r},\,\lambda)}$ is defined as 
\begin{equation}
m({\bf q})\,=\,\frac{1}{n}\sum_{k\,=\,1}^{n}\,{\bf q}^k\,,
\end{equation}
and the standard deviation of ${{\bf q}(\psi,\,{\bf r},\,\lambda)}$ is defined as
\begin{align}
&\sigma[{\bf q}(\psi,\,{\bf r},\,\lambda)]\, \notag \\
&:=\,\sqrt{\frac{1}{n}\sum_{k\,=\,1}^{n}\,\left\{\,{\bf q}^k(\psi)\,-\,m({\bf q})\right\}\,
\left\{\,{\bf q}^k(2\pi-\psi)\,-\,m({\bf q})\right\}^{\dagger}\;}.\label{standipo}
\end{align}

Note that in this definition ${{\bf q}(\psi)}$ is
coordinated by ${\psi}$ to rotate from ${0}$ to ${2\pi}$, whereas the conjugate ${{\bf q}^{\dagger}(2\pi-\psi)}$ is
coordinated by ${\psi}$ to rotate from ${2\pi}$ to ${0}$. Thus, for a given value of ${\lambda}$, both ${{\bf q}(\psi)}$
and ${{\bf q}^{\dagger}(2\pi-\psi)}$ represent the same sense of rotation about ${\bf r}$ (either both represent clockwise
rotations or both represent counterclockwise rotations). This is crucial for the calculation of standard deviation, for
it is supposed to give the average rotational distance within ${S^3}$ from its mean, with the average being taken, not over
rotational distances within a fixed orientation of ${S^3}$, but over the changes in the orientation ${\lambda}$ of ${S^3}$
itself. Note also that, according to the definition (\ref{quat-0}),
${{\bf q}(\psi)}$ and its conjugate ${{\bf q}^{\dagger}(\psi)}$ satisfy the following relation:
\begin{equation}
{\bf q}^{\dagger}(2\pi-\psi)\,=\,-\,{\bf q}(\psi)\,. \label{relat}
\end{equation}
Consequently, the standard deviation of both ${{\bf q}^{\dagger}(2\pi-\psi)}$ and ${-\,{\bf q}(\psi)}$
must necessarily give the same number:
\begin{equation}
\sigma[\,{\bf q}^{\dagger}(2\pi-\psi)]\,\equiv\,\sigma[\,-\,{\bf q}(\psi)]\,.
\end{equation}
It is easy to verify that definition (\ref{standipo}) for the standard deviation of ${{\bf q}(\psi)}$ does indeed satisfy
this requirement, at least when ${m({\bf q})=0}$. What is more, from Eq.${\,}$(\ref{relat}) we note that the quantity
being averaged in the definition (\ref{standipo}) is essentially ${-\,{\bf q}\,{\bf q}}$. This quantity is insensitive to spinorial
sign changes, such as ${{\bf q}\rightarrow -\,{\bf q}}$, but transforms into ${-\,{\bf q}^{\dagger}{\bf q}^{\dagger}}$ under
orientation changes, such as ${{\lambda}\rightarrow -{\lambda}}$. By contrast, the quantity
${-\,{\bf q}\,{\bf q}^{\dagger}}$ would be insensitive to both spinorial sign changes as well as orientation changes.
Thus ${\sigma[{\bf q}(\lambda)]}$, as defined in (\ref{standipo}), is designed to remain sensitive to orientation changes
for correctly computing its averaging function on ${{\bf q}(\lambda)}$ in the present context.

Now, in order to evaluate ${\sigma({\mathscr A})}$ and ${\sigma({\bf L}_{\bf a})}$,
we rewrite the quaternion (\ref{quat-0}) rotating
about ${{\bf r}={\bf a}}$ as a product
\begin{equation}
{\bf q}(\psi,\,{\bf a},\,\lambda)\,=\,
{\bf p}(\psi,\,{\bf a})\,{\bf L}({\bf a},\,\lambda) \label{pro}
\end{equation}
of a non-random, non-pure quaternion
\begin{align}
{\bf p}(\psi,\,{\bf a})\,
&:=\,\sin\left(\frac{\psi}{2}\right)\,-\,{\bf D(a)}\,\cos\left(\frac{\psi}{2}\right) \notag \\
&=\,\exp\left\{-\,{\bf D(a)}\!\left(\frac{\pi-\psi}{2}\right)\!\right\}
\label{genno}
\end{align}
and a random, unit bivector ${{\bf L}({\bf a},\,\lambda)}$ satisfying
\begin{equation}
\frac{1}{n}\sum_{k\,=\,1}^{n}\,
{\bf L}({\bf a},\,\lambda^k)\,{\bf L}^{\dagger}({\bf a},\,\lambda^k)\,=\,1\,. \label{normL}
\end{equation}
Note that ${{\bf p}(\psi,\,{\bf a})}$ reduces to the unit bivector ${\pm\,{\bf D}\!\left({\bf a}\right)}$ for rotation
angles ${\psi=0}$, ${\psi=2\pi}$, and ${\psi=4\pi}$. Moreover, using the relations
${{\bf L}\!\left({\bf a},\,\lambda\right)=\lambda\,{\bf D}\!\left({\bf a}\right)}$ and ${{\bf D}^2({\bf a})=-\,1}$
it can be easily checked that the product in
(\ref{pro}) is indeed equivalent to the quaternion defined in (\ref{quat-0}) for ${{\bf r}={\bf a}}$. It is also easy to check
that the non-random quaternion ${{\bf p}(\psi,\,{\bf a})}$ satisfies the following relation with its conjugate:
\begin{equation}
{\bf p}^{\dagger}(2\pi-\psi,\,{\bf a})\,=\,{\bf p}(\psi,\,{\bf a})\,.
\end{equation}
Consequently we have
\begin{align}
{\bf q}^{\dagger}(2\pi-\psi,\,{\bf a},\,\lambda)\,&=\,\left\{{\bf p}
(2\pi-\psi,\,{\bf a})\,{\bf L}({\bf a},\,\lambda)\right\}^{\dagger} \notag \\
\,&=\,{\bf L}^{\dagger}({\bf a},\,\lambda)\,{\bf p}^{\dagger}(2\pi-\psi,\,{\bf a}) \notag \\
\,&=\,{\bf L}^{\dagger}({\bf a},\,\lambda)\,{\bf p}(\psi,\,{\bf a})\,. \label{non-pro}
\end{align}
Thus, substituting for ${{\bf q}(\psi,\,{\bf a},\,\lambda)}$ and ${{\bf q}^{\dagger}(2\pi-\psi,\,{\bf a},\,\lambda)}$ from
Eqs.${\,}$(\ref{pro}) and (\ref{non-pro}) into Eq.${\,}$(\ref{standipo}), together with
\begin{align}
m({\bf q}_{\bf a}^{\,})\,=\,\frac{1}{n}\sum_{k\,=\,1}^{n}\,{\bf q}_{\bf a}^k
\,&=\,{\bf p}(\psi,\,{\bf a})\,\left\{\frac{1}{n}\sum_{k\,=\,1}^{n}\,{\bf L}({\bf a},\,\lambda^k)\right\} \notag \\
\,&=\,{\bf p}(\psi,\,{\bf a})\,\left\{\frac{1}{n}\sum_{k\,=\,1}^{n}\,\lambda^k\right\}\,{\bf D}({\bf a}) \notag \\
\,&=\,0\,,
\end{align}
we have
\begin{align}
\sigma&[{\bf q}(\psi,\,{\bf a},\,\lambda)]\, \notag \\
&=\sqrt{\frac{1}{n}\sum_{k\,=\,1}^{n}\,\left\{\,{\bf p}(\psi,\,{\bf a})\,{\bf L}({\bf a},\,\lambda^k)\,\right\}\,
\left\{\,{\bf L}^{\dagger}({\bf a},\,\lambda^k)\,{\bf p}(\psi,\,{\bf a})\,\right\}\,} \notag \\
&=\,\sqrt{\,{\bf p}(\psi,\,{\bf a})\,\left\{\,\frac{1}{n}\sum_{k\,=\,1}^{n}\,
{\bf L}({\bf a},\,\lambda^k)\,{\bf L}^{\dagger}({\bf a},\,\lambda^k)\,\right\}\,{\bf p}(\psi,\,{\bf a})\,} \notag \\
&=\,\sqrt{\,{\bf p}(\psi,\,{\bf a})\,{\bf p}(\psi,\,{\bf a})\,}\,=\,\pm\,{\bf p}(\psi,\,{\bf a})\,. \label{immminot}
\end{align}
Here we have used the normalization of ${{\bf L}({\bf a},\,\lambda)}$ as in (\ref{normL}), and the last equality follows from
definition (\ref{1root1}). It can also be deduced from the polar form of the product
\begin{align}
{\bf p}(\psi,\,{\bf a})\,{\bf p}(\psi,\,{\bf a})\,
&=\,\cos\left(\pi-\psi\right)\,-\,{\bf D(a)}\,\sin\left(\pi-\psi\right) \notag \\
&=\,\exp\left\{\,-\,{\bf D(a)}\left(\pi-\psi\right)\right\}\,.
\end{align}

The result for the standard deviation we have arrived at, namely
\begin{equation}
\sigma[\,{\bf q}(\psi,\,{\bf a},\,\lambda)]\,=\,\pm\,{\bf p}(\psi,\,{\bf a})\,, \label{minot}
\end{equation}
is valid for all possible rotation angles ${\psi}$ between the detector bivector
${-\,{\bf D}({\bf a})}$ and the spin bivector ${{\bf L}({\bf a},\,\lambda)}$. For the special
cases when ${\psi=0}$, ${\pi}$, ${2\pi}$, ${3\pi}$, and  ${4\pi}$, it reduces to the following set of standard deviations:
\begin{align}
\sigma[\,{\bf q}(\psi=0,\,{\bf a},\,\lambda)]\,&=\,\sigma({\mathscr A})\,=\,\pm\,{\bf D}({\bf a}) \notag \\
\sigma[\,{\bf q}(\psi=\pi,\,{\bf a},\,\lambda)]\,&=\,\sigma({\bf L}_{\bf a})\,=\,\pm\,1 \notag \\
\sigma[\,{\bf q}(\psi=2\pi,\,{\bf a},\,\lambda)]\,&=\,\sigma({\mathscr A})\,=\,\pm\,{\bf D}({\bf a}) \notag \\
\sigma[\,{\bf q}(\psi=3\pi,\,{\bf a},\,\lambda)]\,&=\,\sigma({\bf L}_{\bf a})\,=\,\pm\,1 \notag \\
\text{and}\;\;\;\sigma[\,{\bf q}(\psi=4\pi,\,{\bf a},\,\lambda)]\,&=\,\sigma({\mathscr A})
\,=\,\pm\,{\bf D}({\bf a})\,. \label{setresto}
\end{align}
To understand the physical significance of these results, let us first consider the special case when ${\psi=\pi}$. Then
\begin{equation}
{\bf q}(\psi=\pi,\,{\bf a},\,\lambda)\,=\,+\,{\bf L}({\bf a},\,\lambda)\,,
\end{equation}
which can be seen as such from the definition (\ref{quat-0}) above. Similarly, for the conjugate of
${{\bf q}(\psi=\pi,\,{\bf a},\,\lambda)}$ we have
\begin{equation}
{\bf q}^{\dagger}(\psi=\pi,\,{\bf a},\,\lambda)\,=\,-\,{\bf L}({\bf a},\,\lambda)
\,=\,+\,{\bf L}^{\dagger}({\bf a},\,\lambda)\,.
\end{equation}
Moreover, we have ${m({\bf L}_{\bf a})=0}$,
since ${{\bf L}\!\left({\bf a},\,\lambda\right)=+\,\lambda\,{\bf D}\!\left({\bf a}\right)}$
with ${\lambda=\pm\,1}$ being a fair coin. Substituting these results into definition (\ref{standipo})---together
with ${\psi=\pi}$---we arrive at
\begin{equation}
\sigma({\bf L}_{\bf a})\,=\,\sqrt{\frac{1}{n}\sum_{k\,=\,1}^{n}\,
{\bf L}({\bf a},\,\lambda^k)\,{\bf L}^{\dagger}({\bf a},\,\lambda^k)\;}\,=\,\pm\,1\,,
\end{equation}
since ${{\bf L}({\bf a},\,\lambda)\,{\bf L}^{\dagger}({\bf a},\,\lambda)=1}$.
Similarly, we can consider the case when ${\psi=3\pi}$ and again arrive at ${\sigma({\bf L}_{\bf a})=\pm\,1}$.

Next, we consider the three remaining special cases, namely ${\psi=0}$, ${2\pi}$, or ${4\pi}$. These cases correspond to
the measurement results, as defined, for example, in (\ref{88-oi}). To confirm this, recall that a measurement
result such as ${{\mathscr A}({\bf a},\,\lambda)=\pm1}$ is a limiting case of the quaternion${\;}$(\ref{quat-0}):
\begin{align}
\!\!\!{\mathscr A}({\bf a},\,\lambda)\,
&=\,\lim_{{\bf a'}\rightarrow\,{\bf a}}\,{\mathscr A}({\bf a},\,{\bf a'},\,\lambda)\, \notag \\
&=\,\lim_{{\bf a'}\rightarrow\,{\bf a}}\,\left\{\,-\,{\bf D}({\bf a})\,{\bf L}({\bf a'},\,\lambda)\,\right\}\, \notag \\
&=\,\lim_{{\bf a'}\rightarrow\,{\bf a}}\,\left\{\,(\,-\,I\cdot{\bf a})(\,\lambda\,I\cdot{\bf a'})\,\right\} \notag \\
&=\,\lim_{{\bf a'}\rightarrow\,{\bf a}}\,\left\{\,\lambda\,{\bf a}\cdot{\bf a'}
                                   \,+\,\lambda\,I\cdot({\bf a}\times{\bf a'})\,\right\} \notag \\
&=\,\lim_{{\bf a'}\rightarrow\,{\bf a}}\,
                     \left\{\lambda\,\cos\frac{\psi}{2}\,+\,{\bf L}({\bf c},\,\lambda)\,\sin\frac{\psi}{2}\,\right\} \notag \\
&=\,\lim_{{\bf a'}\rightarrow\,{\bf a}}\,\left\{\,{\bf q}(\psi,\,{\bf c},\,\lambda)\,\right\}, \label{amanda}
\end{align}
where ${\psi=2\,\eta_{{\bf a}\,{\bf a'}}}$ is the rotation angle about the axis
${{\bf c}:={\bf a}\times{\bf a'}/|{\bf a}\times{\bf a'}|,\,}$ and ${\eta_{{\bf a}\,{\bf a'}}}$ is the angle
between ${\bf a}$ and ${\bf a'}$. If we now rotate the bivector ${{\bf L}({\bf c},\,\lambda)}$ to ${{\bf L}({\bf a},\,\lambda)}$ as
\begin{equation}
{\bf D}({\bf r})\,{\bf L}({\bf c},\,\lambda)\,{\bf D}^{\dagger}({\bf r})\,=\,{\bf L}({\bf a},\,\lambda)
\end{equation}
using some ${{\bf D}({\bf r})}$, and multiply Eq.{\,}(\ref{amanda}) from the left by ${{\bf D}({\bf r})}$ and
from the right by ${{\bf D}^{\dagger}({\bf r})}$, then we arrive at
\begin{align}
\!\!\!{\mathscr A}({\bf a},\,\lambda)\,
&=\,\lim_{{\bf a'}\rightarrow\,{\bf a}}\,\left\{{\bf D}({\bf r})\,{\bf q}(\psi,\,{\bf c},\,\lambda)\,
{\bf D}^{\dagger}({\bf r})\right\} \notag \\
&=\,\lim_{{\psi}\rightarrow\,{2\kappa\pi}}\,
                     \left\{\lambda\,\cos\frac{\psi}{2}\,+\,{\bf L}({\bf a},\,\lambda)\,\sin\frac{\psi}{2}\,\right\} \notag \\
&=\,\lim_{{\psi}\rightarrow\,{2\kappa\pi}}\,\left\{\,{\bf q}(\psi,\,{\bf a},\,\lambda)\,\right\} \notag \\
&=\,\lim_{{\psi}\rightarrow\,{2\kappa\pi}}\,\left\{\,{\bf p}(\psi,\,{\bf a})\,{\bf L}({\bf a},\,\lambda)\,\right\}, \label{knox}
\end{align}
where ${{\bf p}(\psi,\,{\bf a})}$ is defined in Eq.${\,}$(\ref{genno}).
The limit ${{\bf a'}\rightarrow\,{\bf a}}$ is thus physically equivalent to the limit
${\psi\rightarrow\,2\kappa\pi}$ for ${\kappa\,=\,0,\,1,\;\text{or}\,\;2}$.
We therefore have the following relation between 
${{\mathscr A}({\bf a},\,\lambda)}$ and ${{\bf q}(\psi,\,{\bf a},\,\lambda)}$:
\begin{align}
{\bf q}(\psi=2\kappa\pi,\,{\bf a},\,\lambda)\,
&=\,\pm\;{\bf D}({\bf a})\,{\bf L}({\bf a},\,\lambda)\, \notag \\
&=\,\pm\,{\mathscr A}({\bf a},\,\lambda)\,,
\end{align}
and similarly between ${{\mathscr A}^{\dagger}({\bf a},\,\lambda)}$ and ${{\bf q}^{\dagger}(\psi,\,{\bf a},\,\lambda)}$:
\begin{align}
{\bf q}^{\dagger}(\psi=2\kappa\pi,\,{\bf a},\,\lambda)
\,&=\,\left\{\,\pm\,{\bf D}({\bf a})\,{\bf L}({\bf a},\,\lambda)\,\right\}^{\dagger} \notag \\
\,&=\,\pm\,{\mathscr A}^{\dagger}({\bf a},\,\lambda)\,. \label{15-2}
\end{align}
For example, for ${\psi=0}$ the definition (\ref{quat-0}) leads to
\begin{equation}
{\bf q}(\psi=0,\,{\bf a},\,\lambda)\,=\,-\,{\bf D}({\bf a})\,{\bf L}({\bf a},\,\lambda)
\,=\,+\,{\mathscr A}({\bf a},\,\lambda)\,. \label{14}
\end{equation}
This tells us that in the ${\psi\rightarrow\,0}$ limit the quaternion ${{\bf q}(\psi,\,{\bf a},\,\lambda)}$ reduces to the
scalar point ${\,-\,{\bf D}({\bf a})\,{\bf L}({\bf a},\,\lambda)}$ of ${S^3}$. Moreover, we have ${m({\mathscr A})=0}$, since
${m({\bf L}_{\bf a})=0}$
as we saw above. On the other hand, from the definition (\ref{quat-0}) of ${{\bf q}(\psi,\,{\bf a},\,\lambda)}$
we also have the following relation between the
conjugate variables ${{\mathscr A}^{\dagger}({\bf a},\,\lambda)}$ and ${{\bf q}^{\dagger}(\psi=2\pi,\,{\bf a},\,\lambda)}$:
\begin{align}
{\bf q}^{\dagger}(\psi=2\pi,\,{\bf a},\,\lambda)
\,&=\,+\,\left\{\,{\bf D}({\bf a})\,{\bf L}({\bf a},\,\lambda)\,\right\}^{\dagger} \notag \\
\,&=\,+\,{\bf L}^{\dagger}({\bf a},\,\lambda)\,{\bf D}^{\dagger}({\bf a}) \notag \\
\,&=\,-\,{\bf L}^{\dagger}({\bf a},\,\lambda)\,{\bf D}({\bf a}) \notag \\
\,&=\,-\,{\mathscr A}^{\dagger}({\bf a},\,\lambda)\,. \label{15-1}
\end{align}
This tells us that in the ${\psi\rightarrow\,2\,\pi}$ limit the quaternion ${{\bf q}^{\dagger}(\psi,\,{\bf a},\,\lambda)}$
reduces to the scalar point ${\,-\,{\bf L}^{\dagger}({\bf a},\,\lambda)\,{\bf D}({\bf a})}$ of ${S^3}$.
Thus the case ${\psi=0}$ does indeed correspond to the measurement events. The physical significance of the two remaining
cases, namely ${\psi=2\pi}$ and ${4\pi}$, can be verified similarly, confirming the set of results listed in (\ref{setresto}):
\begin{equation}
\sigma({\mathscr A})\,=\,\pm\,{\bf D}({\bf a})\,. \label{a30}
\end{equation}
Substituting this into Eq.${\,}$(\ref{var-a}) then immediately leads to the standard scores:
\begin{align}
{A}({\bf a},\,\lambda)
\,&=\frac{\,\pm\,{\mathscr A}({\bf a},\,{\lambda})\,-\,
{\overline{{\mathscr A}({\bf a},\,{\lambda})}}}{\sigma({\mathscr A})} \notag \\
\,&=\frac{\,\pm\;{\bf D}({\bf a})\,{\bf L}({\bf a},\,\lambda)\,-\,0\,}{\sigma({\mathscr A})} \notag \\
\,&=\,\left\{\frac{\,\pm\,{\bf D}({\bf a})\,}{\sigma({\mathscr A})}\right\}\,{\bf L}({\bf a},\,\lambda)
\,=\,{\bf L}({\bf a},\,\lambda)\,. \label{nobust}
\end{align}
This confirms the standard scores derived in Eq.${\,}$(\ref{var-a}).

\begin{figure}
\hrule
\scalebox{0.56}{
\psset{yunit=2cm,xunit=2}
\begin{pspicture}(-0.6,-0.5)(5,5)

\uput[0](-0.73,4.4){{\Large ${\bf q}$}${(f)}$}

\uput[0](-1.64,3.0){\large ${m({\mathscr A})+\sigma({\mathscr A})}$}

\uput[0](-0.87,2.5){\large ${m({\mathscr A})}$}

\uput[0](-1.64,2.0){\large ${m({\mathscr A})-\sigma({\mathscr A})}$}

\uput[-90](5.765,0.15){\Large ${\bf q}$}

\uput[-90](5.55,3.25){\Large ${f}$({\bf S})}

\uput[-90](2.45,-0.1){\large ${m({\bf S})-\sigma({\bf S})}$}

\uput[-90](3.52,-0.1){\large ${m({\bf S})}$}

\uput[-90](4.55,-0.1){\large ${m({\bf S})+\sigma({\bf S})}$}

\pscustom[linestyle=none,fillstyle=solid,fillcolor=gray!25]{%
\psGauss[sigma=0.5,mue=3.5,linewidth=1pt]{2.9}{4.1}
      \psline(4.1,0.0)(2.9,0.0)}

\psGauss[sigma=0.5,mue=3.5,linewidth=1pt,linecolor=blue]{2.0}{5.0}

\uput[-90](2.9,0.519){\large ${\bullet}$}

\uput[-90](3.5,0.931){\large ${\bullet}$}

\uput[-90](4.1,0.519){\large ${\bullet}$}

\psline[linewidth=0.2mm,linestyle=dashed]{-}(2.9,-0.07)(2.9,2.0)

\psline[linewidth=0.2mm,linestyle=dashed]{-}(4.1,-0.07)(4.1,3.0)

\begin{rotate}{-90}
\pscustom[linestyle=none,fillstyle=solid,fillcolor=gray!25]{%
\psGauss[sigma=0.5,mue=-2.5,linewidth=1pt,fillstyle=solid,fillcolor=gray!25]{-3.00}{-2.0}%
      \psline(-2.0,0.0)(-3.0,0.0)}
\end{rotate}

\psaxes[labels=none,ticksize=0pt,arrowinset=0.3,arrowsize=3pt 4,arrowlength=3]{->}(0,0)(-0.2,-0.2)(5.5,4.5)

\psline[linewidth=0.2mm,linestyle=dashed]{-}(-0.07,2.0)(2.9,2.0)

\psline[linewidth=0.2mm,linestyle=dashed]{-}(-0.07,3.0)(4.1,3.0)

\begin{rotate}{-90}
\psGauss[sigma=0.5,mue=-2.5,linewidth=1pt,linecolor=red]{-4.0}{-1.0}%
\end{rotate}

\uput[-90](0.48,3.134){\large ${\bullet}$}

\uput[-90](0.8,2.625){\large ${\bullet}$}

\uput[-90](0.48,2.125){\large ${\bullet}$}

\psline[linewidth=0.3mm]{-}(2.0,1.26)(4.75,3.53)

\psline[linewidth=0.2mm,linestyle=dashed]{-}(3.5,-0.07)(3.5,2.5)

\psline[linewidth=0.2mm,linestyle=dashed]{-}(-0.07,2.5)(3.5,2.5)

\pscurve[linewidth=1.5pt,linecolor=green]{-}(1.5,0.75)(1.9,0.8)(3.5,2.483)(4.0,2.79)(4.75,3.015)(5.15,3.05)

\end{pspicture}}
\vspace{0.25cm}
\hrule
\caption{Propagation of error within a parallelized 3-sphere.\break}
\label{fig-88}
\vspace{0.3cm}
\hrule
\end{figure}
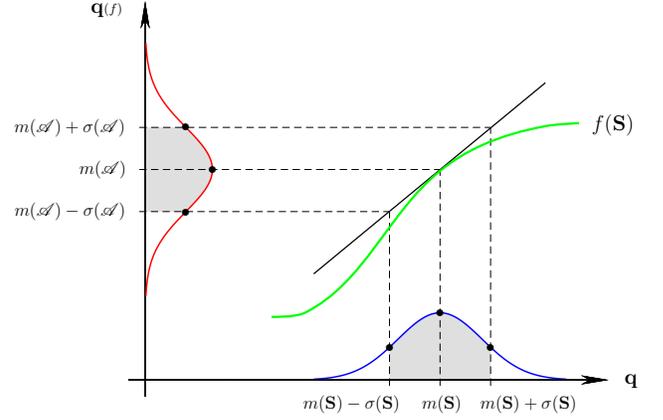

The above derivation of the standard scores holds, in fact, for any general element
${{\bf q}(\psi,\,{\bf a},\,\lambda)}$ of the set ${S^3}$: 
\begin{align}
{\bf {\widehat q}}(\psi,\,{\bf a},\,\lambda)
\,&=\frac{\,{\bf q}(\psi,\,{\bf a},\,\lambda)\,-\,m({\bf q})}{\sigma[\,{\bf q}(\psi,\,{\bf a},\,\lambda)]} \notag \\
\,&=\frac{\,{\bf q}(\psi,\,{\bf a},\,\lambda)\,-\,0\,}{{\bf p}(\psi,\,{\bf a})} \notag \\
\,&=\,{\bf q}(\psi,\,{\bf a},\,\lambda)\,{\bf p}^{\dagger}(\psi,\,{\bf a}) \notag \\
\,&=\,{\bf L}({\bf a},\,\lambda)\,, \label{robust}
\end{align}
where we have used the definitions (\ref{quat-0}) and (\ref{genno}) of ${{\bf q}(\psi,\,{\bf a},\,\lambda)}$
and ${{\bf p}(\psi,\,{\bf a})}$, respectively. It is important to note here that the standard scores
${{\bf L}({\bf a},\,\lambda)}$ so derived are independent of the rotation angle ${\psi}$. This shows that
our derivation of the correlation in Eqs.${\,}$(\ref{corwedid}) through (\ref{hjro}) is quite a {\it robust} result.
The correlation between {\it any} two points ${{\bf q}(\psi_{\bf a},\,{\bf a},\,\lambda)}$ and
${{\bf q}(\psi_{\bf b},\,{\bf b},\,\lambda)}$ within ${S^3}$ is always equal to ${-\,{\bf a}\cdot{\bf b}}$,
with the scalars ${{\mathscr A}({\bf a},\,\lambda)=\pm\,1}$ and ${{\mathscr B}({\bf b},\,\lambda)=\pm\,1}$
given in (\ref{88-oi}) and (\ref{99-oi}) being merely two special cases of
${{\bf q}(\psi_{\bf a},\,{\bf a},\,\lambda)}$ and ${{\bf q}(\psi_{\bf b},\,{\bf b},\,\lambda)}$ in ${S^3}$.

So far we have assumed the randomness in the variables ${{\mathscr A}{\bf (a, \lambda)}}$ and ${{\mathscr B}{\bf (b, \lambda)}}$
to originate
entirely from the initial state ${\lambda}$ representing the orientation of the 3-sphere. In other words, we have assumed
that the local interaction of the detector ${\bf D(a)}$ with the random spin ${\bf L(a, \lambda)}$ does not introduce additional
randomness in the measurement result ${{\mathscr A}{\bf (a, \lambda)}}$. Any realistic interaction between ${\bf D(a)}$ and
${\bf L(a, \lambda)}$, however, would inevitably introduce such a randomness, of purely local, experimental origin. We can model
this randomness by introducing an additional random variable, say $r_{\bf a}\in[\,0,\,1]$, not dependent on ${\lambda}$.
Physically we can think of $r_{\bf a}$ as an alignment parameter between the detector bivector ${\bf D(a)}$ and the spin
bivector ${\bf L(a, \lambda)}$, with the value $r_{\bf a}=1$ representing the perfect alignment. Clearly, introduction of
such an additional parameter will make all the bivectors and quaternions within ${S^3}$ unnormalized.
The corresponding probability density function
(\ref{probdensi-8}) would then represent a Gaussian distribution---provided we assume that the orientation ${\lambda=\pm\,1}$
of $S^3$ itself is distributed non-uniformly between its values $+1$ and $-1$. Moreover, although the individual measurement
results would then fall within the range
\begin{equation}
-\,1\,\leq\,{\mathscr A}({\bf a},\,\lambda)\,\leq\,+\,1\,,
\end{equation}
their mean value would be zero, at least
for a uniformly distributed ${\lambda}$, since the mean value of the product of the independent
random variables ${r_{\bf a}}$ and ${\lambda}$ would then be the product of their mean values:
\begin{equation}
m(r_{\bf a}\,\lambda)\;=\;m(r_{\bf a})\,m(\lambda).
\end{equation}
More importantly, the standard scores computed in the equation (\ref{nobust}) above would not be affected by this more
realistic random process ${r_{\bf a}\,\lambda}$---at least for the special case of uniformly distributed ${\lambda}$---because
these scores involve the ratios of the corresponding raw scores and standard deviations centered about the mean value
${m(r_{\bf a}\,\lambda)=0}$.

Let us now try to understand the propagation of error within ${S^3}$ from this more realistic perspective.
To this end, let the random variable ${{\bf q}(\psi,\,{\bf a},\,\lambda)\in S^3(\lambda)}$ be such that the
measurement results ${{\mathscr A}({\bf a},\,\lambda)\in[-1,\,+1]}$ remain as before, but the bivectors
${{\bf L}({\bf a},\,\lambda)}$ in them are subject to a random process ${r_{\bf a}\,\lambda}$ such that
${{\bf S}({\bf a},\,\lambda,\,r_{\bf a})=r_{\bf a}\,{\bf L}({\bf a},\,\lambda)}$ with ${r_{\bf a}\in[{\hspace{1pt}}0,\,1]}$.
Then the mean value ${m({\bf S})}$ and the standard deviation ${\sigma({\bf S})}$ of ${\bf S}$ would be a bivector and a scalar,
respectively:
\begin{align}
m({\bf S}) & \,=\;\text{a bivector}\;\;\;\;\;\;\;\; \notag \\
\text{and}\;\;\;\sigma({\bf S})& \,=\;\text{a scalar}.
\end{align}
If we now take the detector bivector to be ${{\bf D}({\bf a})=I\cdot{\bf a}}$ as before, then the measurement results can be
identified as ${\,-1 \leq {\mathscr A} = {\bf D}\,{\bf S} \leq +1\,}$ so that ${m({\mathscr A})\geq 0}$. Since ${\bf D}$ is a
non-random bivector, errors generated within ${\mathscr A}$ by the random process ${r_{\bf a}\,\lambda}$ would then stem entirely
from the random bivector ${\bf S}$, and propagate linearly. In other words, the standard deviations within the number
${\mathscr A}$ due to the random process ${r_{\bf a}\,\lambda}$ would then be given by
\begin{equation}
\sigma({\mathscr A})={\bf D}\,\sigma({\bf S}).
\end{equation}
But since ${\sigma({\bf S})}$, as we noted, is a scalar, the typical error ${\sigma({\mathscr A})}$ generated within
${\mathscr A}$ due to the random process ${r_{\bf a}\,\lambda}$ is a bivector. The standardized variable (which must
be used to compare the raw scores ${\mathscr A}$ with other raw scores ${{\mathscr B}\,}$) is thus also a bivector:
${A:={\mathscr A}/{\sigma({\mathscr A})}=\text{scalar}\times{\bf S}}$.

As straightforward as it is, the above conclusion may seem unusual.
It is important to recall, however, that in geometric algebra both scalars and bivectors are treated
on equal footing \cite{Clifford}\cite{Hestenes}. They both behave as real-valued c-numbers, albeit of different grades.
To appreciate the consistency and naturalness of the above conclusion, let
\begin{equation}
{\mathscr A}\,=\,f({\bf S})\,=\,{\bf D}\,{\bf S}\label{defofa-0}
\end{equation}
be a continuous function generated by the geometric product of two bivectors ${{\bf D}({\bf a})}$
and ${{\bf S}({\bf a},\,\lambda,\,r_{\bf a})}$ as before.
The natural question then is: How does a typical error in ${\bf S}$ governed by the probability density
(\ref{probdensi-8})---which can be represented by the 68\% probability interval
\begin{equation}
\left[\,m({\bf S})-\sigma({\bf S}),\;m({\bf S})+\sigma({\bf S})\,\right]\label{int-1no}
\end{equation}
as shown in the Fig.${\,}$\ref{fig-88}---propagate from the random bivector ${\bf S}$ to the random
scalar ${\mathscr A}$, through the function ${f({\bf S}) = {\bf D}\,{\bf S}}$? To answer this question
we note that the two end points of the interval (\ref{int-1no}) represent two points, say
${{\bf q}^-}$ and ${{\bf q}^+}$, of the 3-sphere, which is a Riemannian manifold. The
geometro-algebraic distance between the points ${{\bf q}^-}$ and ${{\bf q}^+}$ can therefore be defined, say, as
\begin{equation}
d{\hspace{-1pt}}\left({\bf q}^-\!,\,{\bf q}^+\right)
\,=\,\left({\bf q}^- -\,{\bf q}^+\right)\times \text{sign}\!\left({\bf q}^- -\,{\bf q}^+\right).
\end{equation}
Moreover, from definition (\ref{defofa-0}) of ${\mathscr A}$ and a first-order Taylor expansion of
the function ${f({\bf S})}$ about the point ${{\bf S} = m({\bf S})}$ we obtain
\begin{equation}
{\mathscr A}\,=\,f(m({\bf S}))\,+\,\frac{\partial f}{\partial {\bf S}}\bigg|_{{\bf S}\,=\;m({\bf S})}
({\bf S}\,-\,m({\bf S}))\,+\,\dots \label{Tylore}
\end{equation}
Now it is evident that the slope ${{\partial f}/{\partial {\bf S}}={\bf D}}$ of this line is a constant.
Therefore the mean ${m({\mathscr A})}$ and the standard deviation ${\sigma({\mathscr A})}$ of the distribution of
${{\mathscr A}\!}$ can be obtained by setting ${{\bf S}=m({\bf S})}$ and ${{\bf S}=\sigma({\bf S})}$:
\begin{align}
m({\mathscr A})\,&=\,f(m({\bf S}))\,=\,{\bf D}\, m({\bf S})\,=\,\text{a scalar} \\
\text{and}\;\;\sigma({\mathscr A})\,&=\,\frac{\partial f}{\partial {\bf S}}\,\sigma({\bf S})
\,=\,{\bf D}\,\sigma({\bf S})\,=\,\text{a bivector}.
\end{align}
The probability distribution of ${\mathscr A}$ is thus represented by the 68\% interval
\begin{equation}
\left[\,m({\mathscr A})-\sigma({\mathscr A}),\;m({\mathscr A})+\sigma({\mathscr A})\,\right].\label{int-2no}
\end{equation}
If we now set ${r_{\bf a}=1}$ and thereby assume that ${\bf S}$ is in fact
the unit bivector ${\bf L}$ with a vanishing mean, then we have ${m({\mathscr A})=0}$
and ${\sigma({\mathscr A})=\pm\,{\bf D}}$, as in equation (\ref{a30}) above.

\makeatletter
\renewcommand*{\rightarrowfill@}{%
   \arrowfill@\relbar\relbar\chemarrow}
\makeatother

Finally, it is instructive to note that, geometrically, the propagation of error within ${S^3}$
is equivalent to a simple change in the perspective (cf. Fig.${\,}$\ref{fig-88}):  
\begin{equation}
S^3\ni\,
  \underbrace{
    \overbrace{m({\bf S})}^\text{bivector} \;\pm\;\,
    \overbrace{\sigma({\bf S})}^\text{scalar}
   }_\text{quaternion}
\;\;\xrightarrow[]{\text{\;\;\;\;\;\;${f({\bf S})}$\;\;\;\;\;}}\;\;
\underbrace{
    \overbrace{m({\mathscr A})}^\text{scalar} \,\;\pm\;
    \overbrace{\sigma({\mathscr A})}^\text{bivector}
   }_\text{quaternion}.
\end{equation}
In particular, the probability density of the scalar ${\mathscr A}$ over ${S^3}$ corresponding to interval (\ref{int-2no}) is
equivalent to that of the bivector ${\bf S}$ over ${S^3}$ corresponding to interval (\ref{int-1no}).

\section{Derivation of Tsirel'son's Bound}

For completeness of our derivation of the correlation (\ref{corwedid}), in this appendix we derive the Tsirel'son's bound
on any possible binary correlation. To this end, consider four observation axes,
${\bf a}$, ${\bf a'}$, ${\bf b}$, and ${\bf b'}$, for the experiment described in section IV.
Then the corresponding CHSH string of expectation values \cite{Christian}, namely the coefficient
\begin{equation}
{\cal E}({\bf a},\,{\bf b})\,+\,{\cal E}({\bf a},\,{\bf b'})\,+\,
{\cal E}({\bf a'},\,{\bf b})\,-\,{\cal E}({\bf a'},\,{\bf b'})\,, \label{B1-11}
\end{equation}
would be bounded by the constant ${2\sqrt{2}}$, as discovered by Tsirel'son within the setting of Clifford algebra applied
to quantum mechanics \cite{Christian}\cite{bounds}. Here each of the joint expectation values of the raw scores
${{\mathscr A}({\bf a},\,{\lambda})=\pm\,1}$ and ${{\mathscr B}({\bf b},\,{\lambda})=\pm\,1}$ are defined as
\begin{equation}
{\cal E}({\bf a},\,{\bf b})\,=\lim_{\,n\,\gg\,1}\left[\frac{1}{n}\sum_{k\,=\,1}^{n}\,
{\mathscr A}({\bf a},\,{\lambda}^k)\;{\mathscr B}({\bf b},\,{\lambda}^k)\right],\label{exppeu}
\end{equation}
with the binary numbers ${{\mathscr A}({\bf a},\,{\lambda})}$ defined by the limit
\begin{align}
S^3\ni\pm\,1\,=\,{\mathscr A}({\bf a},\,\lambda)\,
&=\,\lim_{{\psi}\rightarrow\,{2\kappa\pi}}\,\left\{\,{\bf q}(\psi,\,{\bf a},\,\lambda)\,\right\} \notag \\
&=\,-\,{\bf D}({\bf a})\,{\bf L}({\bf a},\,\lambda).
\end{align}
Thus ${{\mathscr A}({\bf a},\,{\lambda})}$ and  ${{\mathscr B}({\bf b},\,{\lambda})}$ are points of a parallelized 3-sphere
and ${{\cal E}({\bf a},\,{\bf b})}$ evaluated in (\ref{exppeu}) gives correlation between such points of the 3-sphere. The
correct value of the correlation ${{\cal E}({\bf a},\,{\bf b})}$, however, cannot be obtained without first appreciating
the fact that ${{\mathscr A}({\bf a},\,{\lambda})=\pm\,1}$ is a product of a ${\lambda}$-independent constant, namely
${-\,{\bf D}({\bf a})}$, and a ${\lambda}$-dependent variable, namely ${{\bf L}({\bf a},\,\lambda)}$. Thus the correct
value of ${{\cal E}({\bf a},\,{\bf b})}$ is obtained by calculating the covariance of the corresponding standardized variables
\begin{align}
A_{\bf a}({\lambda})\,\equiv\,A({\bf a},\,\lambda)\,&=\,{\bf L}({\bf a},\,\lambda) \label{dumtit-1} \\
\text{and}\;\;\;B_{\bf b}({\lambda})\,\equiv\,B({\bf b},\,\lambda)\,&=\,{\bf L}({\bf b},\,\lambda)\,, \label{dumtit-2}
\end{align}
as we discussed just below Eq.${\,}$(\ref{stan}). In other words, the correlation between the raw scores
${{\mathscr A}({\bf a},\,{\lambda})}$ and ${{\mathscr B}({\bf b},\,{\lambda})}$ is the product moment coefficient
\begin{equation}
{\cal E}({\bf a},\,{\bf b})\,=\lim_{\,n\,\gg\,1}\left[\frac{1}{n}\sum_{k\,=\,1}^{n}\,
{A}({\bf a},\,{\lambda}^k)\;{B}({\bf b},\,{\lambda}^k)\right].\label{stand-exppeu}
\end{equation}
The numerical value of this coefficient is then necessarily equal to the value of the correlation calculated in (\ref{exppeu}). 

Using the above expression for ${{\cal E}({\bf a},\,{\bf b})}$
the Bell-CHSH string of expectation values (\ref{B1-11}) can now be rewritten in terms of
the standard scores ${{A}({\bf a},\,{\lambda})}$ and ${{B}({\bf b},\,{\lambda})}$ as
\begin{align}
\lim_{\,n\,\gg\,1}\Bigg[\frac{1}{n}\sum_{k\,=\,1}^{n}\,&\big\{
A_{\bf a}({\lambda}^k)\,B_{\bf b}({\lambda}^k)\,+\,
A_{\bf a}({\lambda}^k)\,B_{\bf b'}({\lambda}^k)\, \notag \\
&+\, A_{\bf a'}({\lambda}^k)\,B_{\bf b}({\lambda}^k)\,-\,
A_{\bf a'}({\lambda}^k)\,B_{\bf b'}({\lambda}^k)\big\}\Bigg]. \label{probnonint}
\end{align}
But since ${A_{\bf a}({\lambda})={\bf L}({\bf a},\,\lambda)}$ and ${B_{\bf b}({\lambda})={\bf L}({\bf b},\,\lambda)}$ are
two independent equatorial points of ${S^3}$, we can take them to belong to two disconnected ``sections'' of ${S^3}$
({\it i.e.}, two disconnected 2-spheres within ${S^3}$), satisfying
\begin{equation}
\left[\,A_{\bf n}({\lambda}),\,B_{\bf n'}({\lambda})\,\right]\,=\,0\,
\;\;\;\forall\;\,{\bf n}\;\,{\rm and}\;\,{\bf n'}\,\in\,{\rm I\!R}^3,\label{com}
\end{equation}
which is equivalent to anticipating a null outcome along the direction ${{\bf n}\times{\bf n'}}$ exclusive
to both ${\bf n}$ and ${\bf n'}$.
If we now square the integrand of equation (\ref{probnonint}), use the above commutation relations, and use the fact
that all bivectors square to ${-1}$, then the absolute value of the Bell-CHSH string (\ref{B1-11}) leads to the
following variance inequality \cite{Christian}:
\begin{align}
|{\cal E}({\bf a},&\,{\bf b})\,+\,{\cal E}({\bf a},\,{\bf b'})\,+\,
{\cal E}({\bf a'},\,{\bf b})\,-\,{\cal E}({\bf a'},\,{\bf b'})|\, \notag \\
&\leqslant\sqrt{\lim_{\,n\,\gg\,1}\left[\frac{1}{n}\sum_{k\,=\,1}^{n}\,
\big\{\,4\,+\,4\,{\mathscr T}_{\,{\bf a\,a'}}({\lambda}^k)\,{\mathscr T}_{\,{\bf b'\,b}}({\lambda}^k)\,\big\}\right]},\label{yever}
\end{align}
where the classical commutators
\begin{equation}
{\mathscr T}_{\,{\bf a\,a'}}(\lambda):=\frac{1}{2}\left[\,A_{\bf a}(\lambda),\,A_{\bf a'}(\lambda)\right]
\,=\,-\,A_{{\bf a}\times{\bf a'}}(\lambda) \label{aa-potorsion-666}
\end{equation}
and
\begin{equation}
{\mathscr T}_{\,{\bf b'\,b}}(\lambda)
:=\frac{1}{2}\left[\,B_{\bf b'}(\lambda),\,B_{\bf b}(\lambda)\right]\,=\,-\,B_{{\bf b'}\times{\bf b}}(\lambda)\label{bb-potor}
\end{equation}
are the geometric measures of the torsion within ${S^3}$ \cite{Christian}. Thus, it is the non-vanishing torsion
${\mathscr T}$ within ${S^3}$ (or within ${{\rm I\!R}{\rm P}^3}$)---the parallelizing torsion which makes its Riemann curvature
vanish---that is responsible for the stronger-than-linear correlation. We can see this from Eq.${\,}$(\ref{yever}) by setting
${{\mathscr T}=0}$, and in more detail as follows.

Using definitions (\ref{dumtit-1}) and (\ref{dumtit-2}) for ${A_{\bf a}({\lambda})}$ and
${B_{\bf b}({\lambda})}$ and making a repeated use of the bivector identity
\begin{equation}
{\bf L}({\bf a},\,\lambda)\,{\bf L}({\bf a'},\,\lambda)\,=\,-\,{\bf a}\cdot{\bf a'}\,-\,
{\bf L}({\bf a}\times{\bf a'},\,\lambda)\,,
\end{equation}
the above inequality can be further simplified to
\begin{align}
&|{\cal E}({\bf a},\,{\bf b})\,+\,{\cal E}({\bf a},\,{\bf b'})\,+\,
{\cal E}({\bf a'},\,{\bf b})\,-\,{\cal E}({\bf a'},\,{\bf b'})|\, \notag \\
&\leqslant\sqrt{\!4-4\,({{\bf a}}\times{{\bf a}'})\cdot({{\bf b}'}\times{{\bf b}})-
4\!\lim_{\,n\,\gg\,1}\left[\frac{1}{n}\sum_{k\,=\,1}^{n}{{\bf L}}({\bf z},\,\lambda^k)\right]} \notag \\
&\leqslant\sqrt{\!4-4\,({{\bf a}}\times{\bf a'})\cdot({\bf b'}\times{{\bf b}})-
4\!\lim_{\,n\,\gg\,1}\left[\frac{1}{n}\sum_{k\,=\,1}^{n}\lambda^k\right]{{\bf D}}({\bf z})} \notag \\
&\leqslant\,2\,\sqrt{\,1-({{\bf a}}\times{\bf a'})
\cdot({\bf b'}\times{{\bf b}})\,-\,0\,}\,,\label{before-opppo-666}
\end{align}
where ${{\bf z}=({\bf a}\times{\bf a'})\times({\bf b'}\times{\bf b})}$, and---as before---we have used the relation (\ref{OJS})
between ${{\bf L}({\bf z},\,\lambda)}$ and ${{\bf D}({\bf z})}$ from section V. Finally, by noticing that trigonometry dictates
\begin{equation}
-1\leqslant\,({\bf a}\times{\bf a'})\cdot({\bf b'}\times{\bf b})\,\leqslant +1\,,
\end{equation}
the above inequality can be reduced to the form
\begin{equation}
\left|\,{\cal E}({\bf a},\,{\bf b})\,+\,{\cal E}({\bf a},\,{\bf b'})\,+\,
{\cal E}({\bf a'},\,{\bf b})\,-\,{\cal E}({\bf a'},\,{\bf b'})\,\right|\,\leqslant\,2\sqrt{2}\,,
\label{My-CHSH}
\end{equation}
exhibiting the upper bound on all possible correlations.

\renewcommand{\bibnumfmt}[1]{\textrm{[#1]}}

\end{document}